\documentclass[11pt]{article}
  
\usepackage{amsmath,amssymb}

\usepackage{epsfig}  
\usepackage{graphicx}               % Standard graphics package  
\usepackage{url}
\usepackage{hyperref}

\newcommand{\nc}{\newcommand}  

\nc{\beq}{\begin{equation}}  
\nc{\eeq}{\end{equation}}  
\nc{\beqa}{\begin{eqnarray}}  
\nc{\eeqa}{\end{eqnarray}}  
\nc{\bea}{\begin{eqnarray}}  
\nc{\eea}{\end{eqnarray}}  
\nc{\ra}{\rightarrow}  
\nc{\lsim}{\begin{array}{c}\,\sim\vspace{-21pt}\\< \end{array}}  
\nc{\gsim}{\begin{array}{c}\sim\vspace{-21pt}\\> \end{array}}  
\nc{\Tr}{{\rm Tr}}
\nc{\slsh}{\slash\hspace*{-0.22cm}}

\def\be{\begin{equation}}
\def\ee{\end{equation}}
\def\bea{\begin{eqnarray}}
\def\eea{\end{eqnarray}}
\def\bit{\begin{itemize}}
\def\eit{\end{itemize}}
\def\afb{A_{FB}}

\newcommand{\gev}{{\rm GeV}}

\def\to{\rightarrow}

\title{  
\vspace*{-2.3cm}  
\begin{flushright}  
\normalsize{  
 SLAC-PUB-14478
  }  
\end{flushright}  
\vspace{1.5cm}  
\Large  
\textbf{
Improving the Top Quark Forward-Backward Asymmetry Measurement at the LHC
 \\
}\vspace*{1.0cm}   
}

\author{Yang Bai$^{a}$ and Zhenyu Han$^{b}$
\vspace{5mm}
\\
$^{a}$ \normalsize\emph{SLAC National Accelerator Laboratory, 2575 Sand Hill Road, Menlo Park, CA 94025, USA} \\ [2mm]
$^{b}$ \normalsize\emph{Center for the Fundamental Laws of Nature, Harvard University, Cambridge, MA 02138, USA} 
}

\date{\today}

\begin{document}  
\setcounter{page}{0}  
\maketitle  

\vspace*{1cm}  
\begin{abstract}
At the LHC, top quark pairs are dominantly produced from gluons, making it difficult to measure the top quark forward-backward asymmetry. To improve the asymmetry measurement, we study variables that can distinguish between top quarks produced from quarks and those from gluons: the invariant mass of the top pair, the rapidity of the top-antitop system in the lab frame, the rapidity of the top quark in the top-antitop rest frame, the top quark polarization and the top-antitop spin correlation. We combine all the variables in a likelihood discriminant method to separate quark-initiated events from gluon-initiated events. We apply our method on models including G-prime's and W-prime's motivated by the recent observation of a large top quark forward-backward asymmetry at the Tevatron. We have found that the significance of the asymmetry measurement can be improved by 10\% to 30\%. At the same time, the central values of the asymmetry increase by 40\% to 100\%. We have also analytically derived the best spin quantization axes for studying top quark polarization as well as spin-correlation for the new physics models.
\end{abstract}  
\thispagestyle{empty}  
\newpage  
  
\setcounter{page}{1}

\baselineskip18pt   

%%%%%%%%%%%%%%%%%%%%%%%%%%%%%%%%
\section{Introduction}
\label{sec:intro}
The recent results on top quark forward-backward asymmetry measurements at the Tevatron have shown interesting evidence of new physics beyond the standard model (SM). Large asymmetry was observed in both the semi-leptonic decay channel at CDF~\cite{AttCDF} and D0~\cite{Abazov:2007qb}, and the di-lepton decay channel at CDF~\cite{CDFAFBdilepton}. While many new physics models have been introduced to explain the top quark forward-backward asymmetry~\cite{ Ferrario:2009bz, Frampton:2009rk, Shu:2009xf, Cao:2010zb, Chivukula:2010fk, Bai:2011ed, Berger:2011ua, Barger:2011ih, Bhattacherjee:2011nr, Blum:2011up,  Grinstein:2011yv, Ligeti:2011vt, Gresham:2011pa, Nelson:2011us, Barcelo:2011fw, Haisch:2011up, Cui:2011xy}, less attention has been paid to directly measuring the top quark forward-backward asymmetry $A_{FB}$ at the LHC. An obvious obstacle is that the LHC is a proton-proton machine as opposed to the Tevatron, which is a proton-anti-proton machine. There is not a universally forward direction for the top quark. This obstacle can be overcome by observing that the valence quarks ($u$ or $d$ quarks) are statistically more energetic than the sea quarks ($\bar u$ or $\bar d$ quarks). So, event by event one can still define the forward direction, making it possible to measure the forward-backward asymmetry~\cite{Kuhn:1998jr, Kuhn:1998kw} (see~\cite{Antunano:2007da, Wang:2010tg, Xiao:2011kp, AguilarSaavedra:2011vw, Hewett:2011wz} for studies for the LHC). 

Another obstacle is that the $t\bar t$ pairs are dominantly produced from gluon initial states, which serve as a huge background for the $A_{FB}$ measurement at the LHC. For the 7 TeV LHC, the production cross section of $t \bar t$ from $gg$ in the SM is approximately a factor of 5 (8) larger than from $u\bar u$ ($d \bar d$). Assuming the observed $A_{FB}$ at the Tevatron does stem from top quarks produced from up or down quarks, as suggested in many beyond-the-SM models, we see that the measurement of $A_{FB}$ at the LHC will be diluted by this additional SM background. Although many new physics explanations of $A_{FB}$ will be tested indirectly by looking for new resonances in $t\bar t$ or dijet channels, it is still important to have a direct measurement of $A_{FB}$ especially when those new resonances are too broad to show up in the standard ``bump" searches. 

In this paper, we discuss how to improve the $A_{FB}$ measurement at the LHC. More specifically, we explore variables that distinguish the $u \bar u \rightarrow t \bar t$ or $d \bar d \rightarrow t \bar t$ production channels from the $g g\rightarrow t \bar t$ channel. Because of the differences in parton distribution functions and differential cross sections, the produced $t \bar t$ system will have different kinematic distributions. For example, the $t\bar t$ invariant mass distribution is often enhanced at large $M_{t\bar t}$ for $u \bar u \rightarrow t \bar t$ or $d \bar d \rightarrow t \bar t$. The rapidity of the $t\bar t$ system in the lab frame, and the top quark rapidity in the $t\bar t$ rest frame also differ for different production mechanisms. 

Other than the simple kinematic variables, we also use the top quark polarization and top-antitop spin correlation information to reduce the $g g\rightarrow t \bar t$ background. In the SM, the leading order QCD production does not generate polarized top quarks, but this is not the case if the new physics explanation of $A_{FB}$ involves new parity-breaking interactions, coupling the left-handed top and the right-handed top differently~\cite{Jezabek:1994zv, Cao:2010nw, Jung:2010yn, Choudhury:2010cd, Krohn:2011tw}. To study the top quark polarization, one first chooses a spin quantization axis and then studies the angles between the top quark decay products and the quantization axis. Hence, it is important to know the spin quantization axis. The traditional wisdom is to use either the beam direction (beam basis) or the top quark moving direction in the $t \bar t$ center-of-mass frame (helicity basis) to quantize the top quark spin. Noticing that neither of those two axes can maximize the top polarization effect, we will derive the ``best quantization axis" for different models and demonstrate its advantages. We emphasize that knowing the ``best quantization axis" is useful not only for cutting off more SM backgrounds and improve the $A_{FB}$ measurement, but also for distinguishing different new physics models. 

Similarly, the spin-correlation between top and anti-top quarks is different for top quark pairs produced from $gg$ and from $q\bar q$. In QCD, when top pairs  are produced near threshold, $t \bar t$ is in a $^3S_1$ state for $q\bar q$ productions and in a $^1S_0$ state for $gg$ productions~\cite{Mahlon:1995zn, Stelzer:1995gc}. As pointed out in Ref.~\cite{Mahlon:1997uc}, one can calculate the ``best quantization axis" to maximize the spin correlation for $q\bar q \rightarrow t \bar t$ in QCD, and the so-called ``off-diagonal basis" depends on the kinematics of the event. We will follow a similar procedure to calculate the ``best axis" in the presence of new physics contribution to $q\bar q \rightarrow t \bar t$, especially for those models without polarized tops. The top quark spin-correlation is an important effect to distinguish the $t \bar t$ resonances as studied in Ref.~\cite{Arai:2007ts, Frederix:2007gi, Bai:2008sk, Baumgart:2011wk}. We believe that the formulas developed in this paper will be useful for identifying the $t\bar t$ resonance properties once the LHC has positive results in those searches.

We combine the variables in a multivariate likelihood method to improve the significance of the asymmetry measurement, defined as  $A_{FB}/\sigma_{A_{FB}}$.  We find that the significance can be improved by 10\% to 30\% for the models we consider. Meanwhile, the central values of the asymmetry increase by 40\% to 100\%, making the measurement less sensitive to systematic uncertainties. 

Our paper is organized as follows. In section~\ref{sec:topFB}, we choose three representative models that explain the $A_{FB}$ results at the Tevatron and calculate their predictions of $A_{FB}$ for the LHC without using additional variables. In section~\ref{sec:basic}, we  consider three basic kinematic variables: the $t\bar t$ invariant mass. $M_{t \bar t}$, the boost of the $t\bar t$ system, $y_{t \bar t}$, and the rapidity of the top quark in the $t\bar t$ rest frame, $y_t$. We study differences between $gg\rightarrow t\bar t$ and $q \bar q \rightarrow t \bar t$ in terms of those variables. We then study top polarization in section~\ref{sec:polarization} and top-antitop spin-correlation in section~\ref{sec:spincorrelation}, with more details given in appendix~\ref{sec:formulas}. In section~\ref{sec:combined}, we demonstrate the combined improvements using all the variables. We then discuss various effects such as experimental cuts and event reconstructions on the improvements when performing a realistic analysis, and conclude our paper in section~\ref{sec:conclusion}.

%%%%%%%%%%%%%%%%%%%%%%%%%%%%%%%%
\section{Top quark forward-backward asymmetry at the LHC}
\label{sec:topFB}
To explain the large top quark forward-backward asymmetry measured at the Tevatron, there are two basic top quark pair production mechanisms. One is through new particle exchange in the $s$-channel such as the axigluon, $G^\prime$, and the other one is through new particles in the $t$-channel such as the diquark, $Z^\prime$ or $W^\prime$. Noticing that $Z^\prime$ model generically predicts copious same-sign tops, which is tightly constrained by the recent analysis from CMS~\cite{Collaboration:2011dk}, we choose $G^\prime$ and $W^\prime$ as two representative examples  in this paper. We will also consider effective contact operators obtained by integrating out a very heavy axigluon. $G^\prime$ and $W^\prime$ couple to the SM quarks as
\beqa
{\cal L}_{G^\prime} &=& - G^{\prime\, a}_\mu \, \left[ \bar{u}\, (g_V^q \gamma^\mu t^a + g_A^q \gamma^\mu\gamma^5 t^a) u \,+\,   \bar{t}\, (g_V^t \gamma^\mu t^a + g_A^t \gamma^\mu\gamma^5 t^a) t \right] \,+\,\cdots \,, \\
{\cal L}_{W^\prime} &=& - W^{\prime +}_\mu\,\bar{t} (g_V \gamma^\mu + g_A \gamma^\mu\gamma^5) d  \,+\,h.c.  \,+\,\cdots \,,
\eeqa 
where we only write down couplings relevant to $A_{FB}$ and $t^a$ is the $SU(3)_c$ generator. The differential production cross section as a function of the top quark production angle is given in~\cite{Hewett:1988xc,Ferrario:2009bz} for the axigluon case and in~\cite{Cheung:2009ch} for the $W^\prime$ case. The measured $A_{FB}$ at the parton level is~\cite{AttCDF} 
\beqa
&&A_{FB}(M_{t\bar t} < 450~\mbox{GeV})= -0.116\pm 0.153\,, \quad\quad 
A_{FB}(M_{t\bar t} \ge 450~\mbox{GeV})= 0.475\pm0.114\,,  \\
&&A_{FB}(|\Delta y| < 1.0)= 0.026\pm 0.118\,, \hspace{2.0cm}
A_{FB}(|\Delta y| \ge 1.0)= 0.611\pm0.256\,.
\eeqa
To fit those experimental data, we consider the following three different model points:
\bit
\item Model A: an axigluon model with $M_{G^\prime} = 2.0$~TeV, $g_A^q = 2.2$, $g_A^t = - 3.2$, $g_V^t=1.0$ and $g_V^q = 0$. Here, ``$q$" represents the first four light quarks. The width is $\Gamma_{G^\prime} = \alpha_s/6(4 g_A^{q\, 2} + 4 g_V^{q\, 2}  + 2 g_A^{t\, 2}  + 2 g_V^{t\, 2}) \,M_{G^\prime} \approx 1.5$~TeV. The predictions for $A_{FB}$  are $(0.10, 0.31)$ for the two invariant mass bins and $(0.12, 0.40)$ for the two rapidity difference bins.  
\item Model B: a $W^\prime$ model with $M_{W^\prime}=400$~GeV and $g_V=g_A=0.9$ (or $g_L=0$ and $g_R = 1.8$). The predictions for $A_{FB}$  are $(0.12, 0.41)$ for the two invariant mass bins and $(0.14, 0.52)$ for the two rapidity difference bins. 
\item Model C: the contact interaction obtained by integrating out a very heavy axigluon (above the center of mass energy of the LHC and the Tevatron), $\xi\,\bar{u} \gamma_\mu \gamma^5t^{a} u \,\bar{t} \gamma^\mu \gamma_5t^{a} t /\Lambda^2$,  with $g_{V}^{q, t} =0$, $\xi= -1$ and $\Lambda \equiv M_{G^\prime}/(g_A^q g_A^t)^{1/2} = 650~\mbox{GeV}$.  The predictions for $A_{FB}$ are $(0.19, 0.53)$ and $(0.23, 0.67)$ for those four bin data. 
\eit
All of the above models satisfy various constraints such as those from  dijet resonance, dijet contact interaction and $t\bar t$ resonance searches at Tevatron~\cite{Bai:2011ed}, though they receive more stringent constraints from the latest LHC results on $t\bar t$ resonance searches \cite{Atlasttbar}. Model A has an axigluon coupling to the top quark with both vector and axial-vector couplings, so the parity is broken. As we will show later, the top quark will be polarized in this case and we can use the top quark polarization to reduce the $gg$ background and improve the $A_{FB}$ measurement. The axigluon width is very large for this model, so the $t\bar t$ resonance would be difficult to discover. For Model B, we only choose right handed couplings for the $W^\prime$ field, based on the severe constraints from electroweak precision observables if $W^\prime$ mixes with the SM $W$ gauge bosons. The $W^\prime$ width is assumed to be small for this case and neglected in our later analysis. Model C only has parity conserving couplings and will not have polarized top quarks. We use this model as an example to study the top and anti-top spin-correlation. 

The measurement of top quark $A_{FB}$ at the LHC is challenging for two reasons. First, it is a proton-proton collider. Unlike Tevatron with proton-antiproton collisions, there is not a fixed forward direction. However, the valence quarks inside the proton most likely carry a larger energy than the sea quarks. So, event by event, we can define the moving direction of the center-of-mass frame with respect to the lab frame as the positive direction to calculate $A_{FB}$. There is a $\sim20\%$ probability that we misidentify the initial parton moving directions. The second reason is that the main mechanism for $t\bar t$ production is  $gg\rightarrow t\bar t$, which does not contribute to $A_{FB}$. Using Madgraph~\cite{Alwall:2011uj}, we obtain the leading order QCD cross sections as $\sigma(gg \rightarrow t\bar t) \approx 71$~pb, $\sigma(u\bar u \rightarrow t\bar t) \approx 14.5$~pb and $\sigma(d\bar d \rightarrow t\bar t) \approx 8.6$~pb for the 7TeV LHC. One can immediately see that there is an additional $t\bar t$ background from $gg$ by a factor of 5. This is different from the Tevatron case, where  $\sigma(gg \rightarrow t\bar t)$ is a subdominant part in the total production cross section. Including all $t\bar t$ pairs and neglecting other SM backgrounds, we follow the above definition of the forward direction and estimate the $A_{FB}$ measurements for the three models as
\beqa
\mbox{Model A:}&& \quad A_{FB}(M_{t\bar t} > 450~\mbox{GeV}) = 0.046\pm0.015\,, \nonumber \\
\mbox{Model B:}&& \quad A_{FB}(M_{t\bar t} > 450~\mbox{GeV}) = 0.196\pm0.011\,, \nonumber \\
\mbox{Model C:}&& \quad A_{FB}(M_{t\bar t} > 450~\mbox{GeV}) = 0.099\pm0.015\,, \label{nocut}
\eeqa
where, to estimate the statistic errors we have assumed a 3 fb$^{-1}$ luminosity and semileptonic decays for the $t\bar t$ system. Furthermore, we have multiplied the total number of events by a 10\% event acceptance (depending on cuts, the acceptance could be even higher than this value, see Ref.~\cite{Atlasttbar} for example), so there are around $5000$ events in total. From the numbers in Eq.~(\ref{nocut}), one can already see that the early LHC running can measure $A_{FB}$ at a large confidence level\footnote{After imposing the $M_{t\bar t} > 450~\mbox{GeV}$ cut, the production cross section for $gg \rightarrow t\bar t$ is 37.6 pb, the ``signal" production cross sections are 13.1, 46.6 and 17.9 pb for Model A, B and C, respectively.  Model B predicts too many $t\bar t$'s and has already been ruled out by the $M_{t\bar t}$ differential cross section distribution~\cite{Atlasttbar}.  Here, we still include this model to illustrate how to improve the LHC $A_{FB}$ measurement for a $t$-channel model and we will also consider a similar model with a smaller coupling.}.

To improve the measurement of $A_{FB}$ at the LHC, we need to distinguish the two $t\bar t$ production mechanisms: from two gluons or from two light quarks. In the following sections, we will study the differences between various kinematics distributions of $t \bar t$ for those two production mechanisms.  

%%%%%%%%%%%%%%%%%%%%%%%%%%%%%%%%
\section{Basic kinematics: invariant mass, $M_{t \bar t}$, the rapidity of the $t\bar t$ system, $y_{t \bar t}$, and the rapidity of the top quark, $y_t$}
\label{sec:basic}

In this section, we first consider some basic kinematic distributions for events generated from the process $gg\rightarrow t \bar t$ and from the three model points. 

\begin{figure}[th!]
\begin{center}
\vspace*{3mm}
\includegraphics[width=0.46\textwidth]{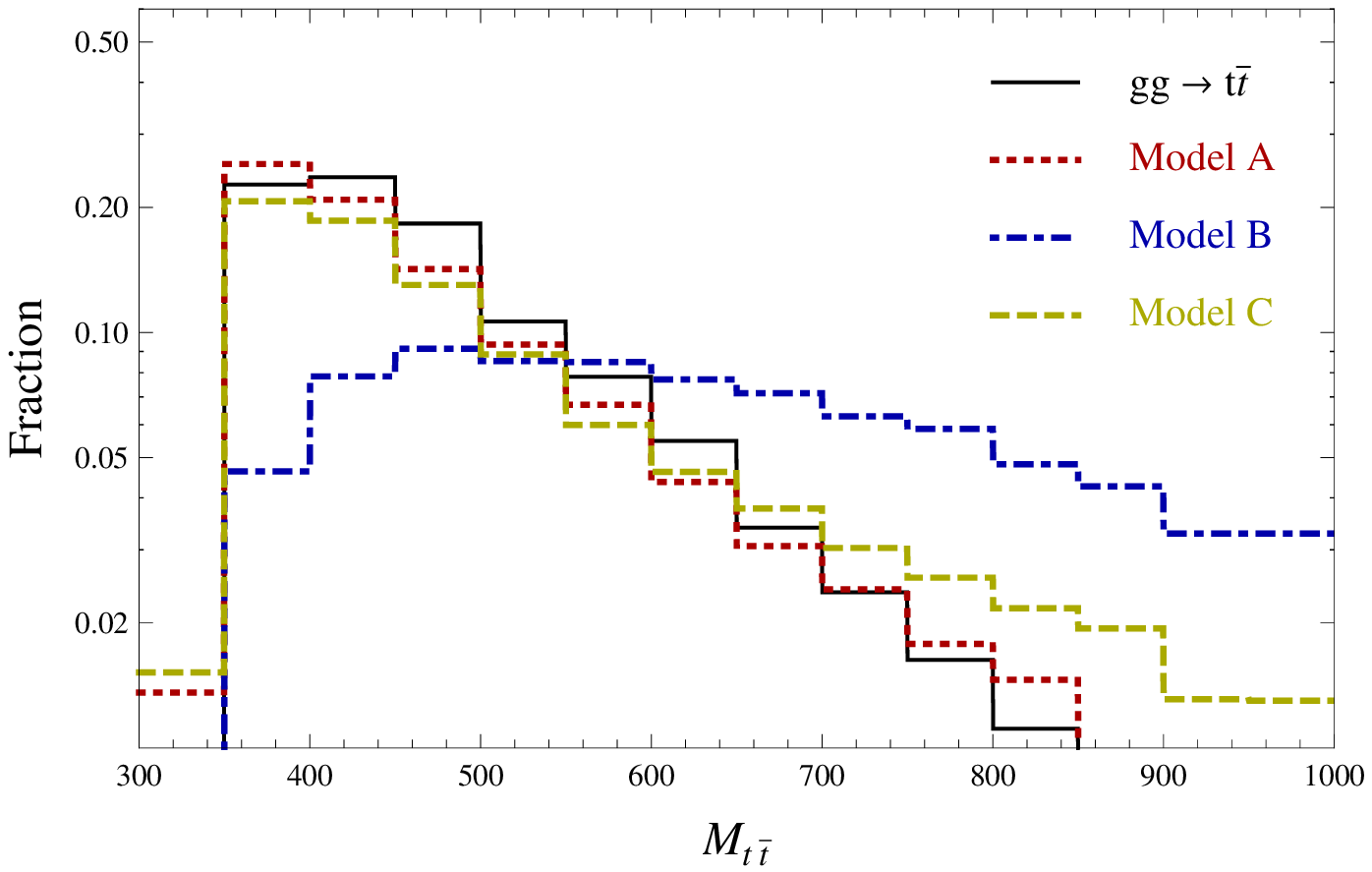}  \quad
\includegraphics[width=0.46\textwidth]{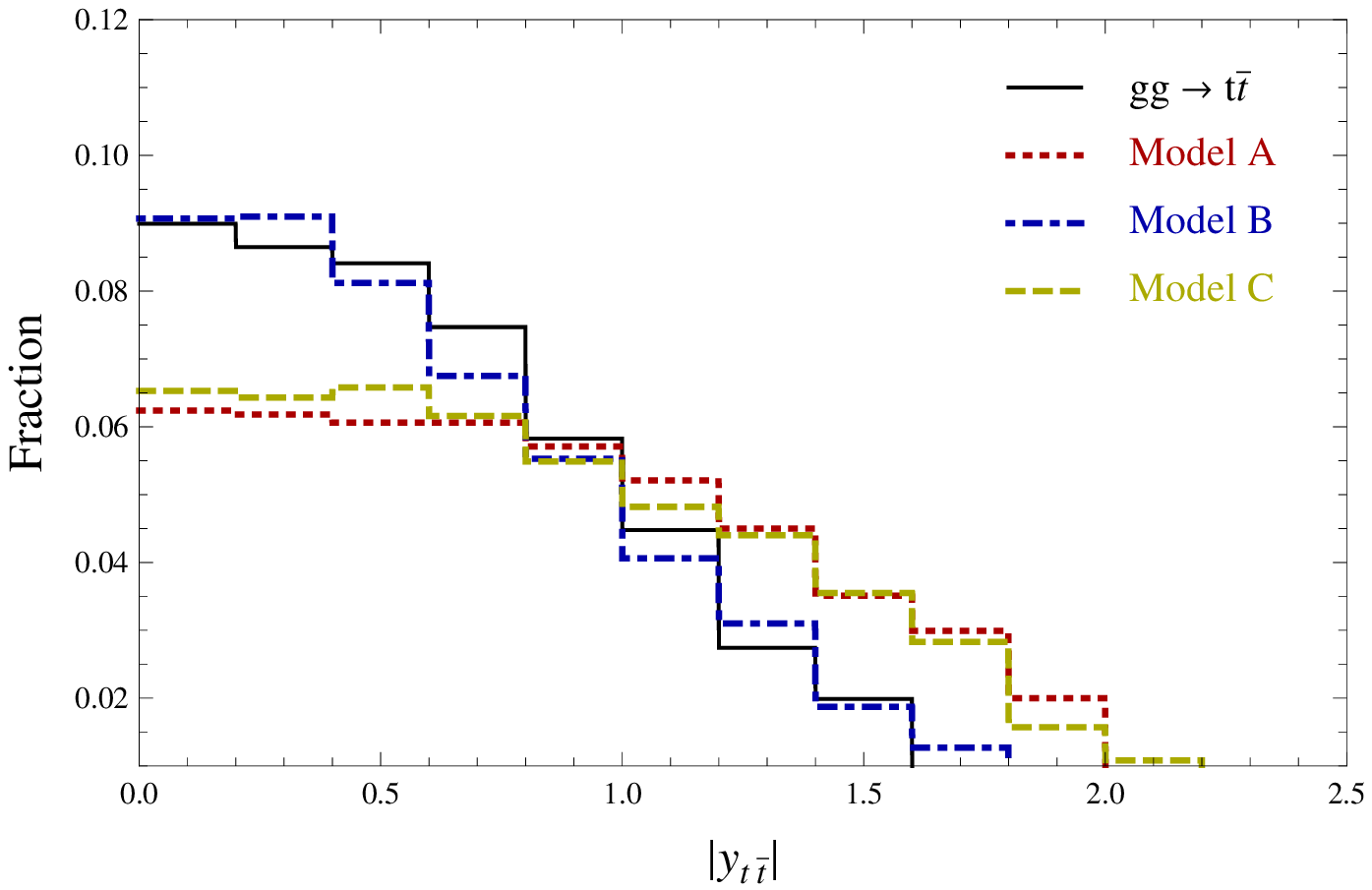} 
\caption{Left panel: the normalized fraction of events as a function of $M_{t\bar t}$. Model A and Model C contains both $u\bar u \rightarrow t \bar t$ and $d\bar d \rightarrow t \bar t$ productions, while Model B only contains $d\bar d \rightarrow t \bar t$. Right panel: the rapidity distributions of the center-of-mass frame of the $t\bar t$ system.}
\label{fig:Pmttrapidity}
\end{center}
\end{figure}

The first variable we consider is the $t\bar t$ invariant mass. Since there are new heavy particles contributing to the top pair production, we anticipate that the tail of the differential cross section in terms of $M_{t\bar t}$ should be lifted. Therefore, the top pairs from new physics should have a harder spectrum than from $gg$. This is illustrated in Fig.~\ref{fig:Pmttrapidity} (left panel). Note that Model B has a much harder spectrum than the other cases. So, we can impose a higher $M_{t \bar t}$ cut to enhance the signal-background ratio. However, as we will show in Section~\ref{sec:combined}, cutting on $M_{t\bar t}$ alone may not increase the statistical significance and may not improve the $\afb$ measurement. Therefore, $M_{t\bar t}$ should be combined with other variables. 

The second variable is the boost of the $t\bar t$ system with respect to the lab frame $y_{t\bar t}$. From the Parton Distribution Functions (PDF's), one expects $|y_{t\bar t}|$ from $u\bar u$ productions to be statistically larger than from $gg$. This is indeed the case as shown in the right panel of Fig.~\ref{fig:Pmttrapidity}, from which one can see that Model A and Model C have more signal events at larger values of $|y_{t\bar t}|$.

Next, we consider the rapidity of the top quark in the $t\bar t$ center-of-mass frame: $|y_t|$. This variable is especially useful for selecting signal events for Model B. This is because of the $t$-channel differential cross section enhancement in the forward direction.
\begin{figure}[th!]
\begin{center}
\vspace*{3mm}
\includegraphics[width=0.46\textwidth]{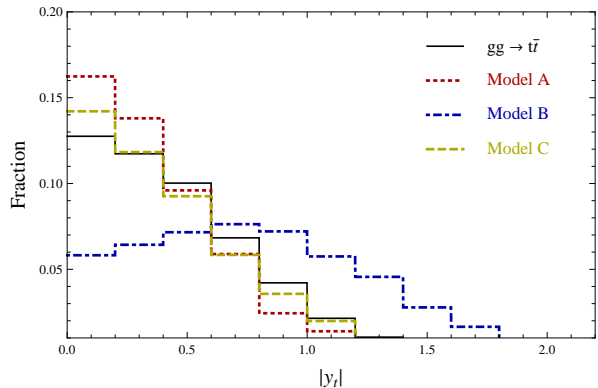} 
\caption{The rapidity distributions of top quark in the center-of-mass frame of the $t\bar t$ system.}
\label{fig:Ptoprapidity}
\end{center}
\end{figure}
From the simulated result in Fig.~\ref{fig:Ptoprapidity}, we see that the $|y_t|$ distribution in Model B (the dot-dashed blue histogram) peaks at around 0.8, which is significantly different from the background (the solid black histogram). On the contrary, for Model A and Model C, the produced top quarks are a little more central but similar to the background. So, we do not anticipate $|y_t|$ as a good variable to improve the $A_{FB}$ measurement for Model A and Model C.

%%%%%%%%%%%%%%%%%%%%%%%%%%%%%%%%
\section{Top quark polarization}
\label{sec:polarization}

The top quark has a short lifetime and decays before hadronization, so its spin information is kept in the angular distributions of the daughter particles. Its polarization as well as spin-correlation with the anti-top quark are different for new physics and the SM. In this section, we utilize the top quark polarization to distinguish signals from the $gg$ background . 

Due to parity conservation, top quarks produced from QCD processes are not polarized. For Model A and Model B, parity is manifestly broken and top quarks generated from new physics are polarized. Choosing a spin-quantization axis for the top quark,
one can study the angular distribution of the daughter particles in the top quark rest frame, with respect to the axis. The differential decay rates for a 100\% polarized top in its rest frame are calculated in Ref.~\cite{Jezabek:1994zv}
\beq
\frac{1}{\Gamma} \frac{d\Gamma}{d\cos{\theta_i} } = \frac{1}{2} ( 1 +  k_i \,\cos{\theta_i}  )\,,
\label{eq:decayparameter}
\eeq
where $\theta_i$ is the angle between the chosen spin-quantization axis and the $i$'th decay product in the top rest frame. For top quark, one has $k_{\ell^+}=k_{\bar d} = k_{\bar s} =1$, $k_{\nu_\ell} = k_u = k_c = -0.31$, $k_{b} = - k_{W^+} = -0.41$. %Those angles $\theta_i$ can only define the probability to determine the top quark is spin-up or spin-down. If $\cos{\theta_i} > y$, then the probability that the top quark has spin up is given by $[2+k_i(1+y)]/4$~\cite{Mahlon:1995zn}. 

Now, the important task is to find the best spin-quantization basis to keep the maximal top polarization information. There are two obvious basis: the beam line basis (the moving direction of the incoming quark in the top rest frame) or the helicity basis (the top quark's moving direction in the $t\bar t$ rest frame). The spin density matrix can be decomposed into independent parts with different spin structures: 
\beq
\rho = A\, {\mathbb{I}_2}\otimes \mathbb{I}_2 + {\bf B}_t \cdot {\bf \sigma} \otimes \mathbb{I}_2 + {\bf B}_{\bar t} \cdot  \mathbb{I}_2 \otimes {\bf \sigma} + C_{ij} \sigma^i \otimes \sigma^j \,,
\label{eq:spindensity}
\eeq
where ${\bf B}_t$ (${\bf B}_{\bar t}$) determines the top (anti-top) quark polarization and $C_{ij}$ denotes the potential spin-correlation between top and anti-top, which is especially useful when ${\bf B}_t= {\bf B}_{\bar t}=0$. The detailed calculations and formulas for those vectors and matrices can be found in Appendix~\ref{sec:formulas}. The polarization vector ${\bf B}_t$ can be decomposed into
\beq
{\bf B}_t = b^{\hat{p}}_t\,{\bf \hat{p}} +  b^{\hat{k}}_t\,{\bf \hat{k}} \,.
\eeq
Here,  $\hat{\bf p} = (0, 0, \pm 1)^T$ is the beam direction and $\hat{\bf k} = (0, s_{\theta^*}, c_{\theta^*})^T$ is the direction of the top quark in the $t\bar t$ rest frame where $\theta^*$ is the top production angle. We have assumed CPT is conserved for the underlying theory, so the projection on the $\bf \hat{p} \times \hat{k}$ direction vanishes. Furthermore, we also assume CP is a good symmetry such that $b_t^{\hat p} = b_{\bar t}^{\hat p}$ and $b_t^{\hat k} = b_{\bar t}^{\hat k}$. Therefore, the anti-top has the same polarization as the top, and we only need to consider the top quark polarization in this section. 

Starting from the axigluon case, Model A, with $g_V^q = 0$ and $g_V^t \neq 0$, we use the following ratio to define the best quantization basis for top quark polarization:
\beq
\frac{b_t^{\hat k}}{b_t^{\hat p}} = \frac{4\,\cos{\theta^*}\,(M^2_{G^\prime}-\hat{s}) (2 m_t -\sqrt{\hat s}) + g_A^t\,g_A^u\,\hat{s}\,\beta\,\left[ \cos{2\theta^*} (\sqrt{\hat s} - 2m_t) - 2m_t + 3\sqrt{\hat{s}} \right]   }
{4\, m_t  (g_A^t g_A^u \,\hat{s}\, \beta\, \cos{\theta^*} - 2 M_{G^\prime}^2  + 2\hat{s} )} \,,
\label{eq:bestModelA}
\eeq
where $\beta = \sqrt{1-4m_t^2/\hat{s}}$. One can see that when $\beta = 0$ or $\sqrt{\hat s} = 2m_t$, the above ratio is zero. This means when two top quarks are produced at rest, the beam line is the best quantization basis. In the other limit when $\hat{s} = M^2_{G^\prime}$ (or when the axigluon is on-shell), this ratio is $(\sec{\theta^*} + \cos{\theta^*})M_{G^\prime}/(2m_t) - \cos{\theta^*}$. So, for $M_{G^\prime} \gg m_t$ the best spin quantization basis is the helicity basis $\bf \hat{k}$. We also note that when $g_V^q = g_V^t = 0$ the top quark is not polarized. 
\begin{figure}[th!]
\begin{center}
\vspace*{3mm}
\includegraphics[width=0.46\textwidth]{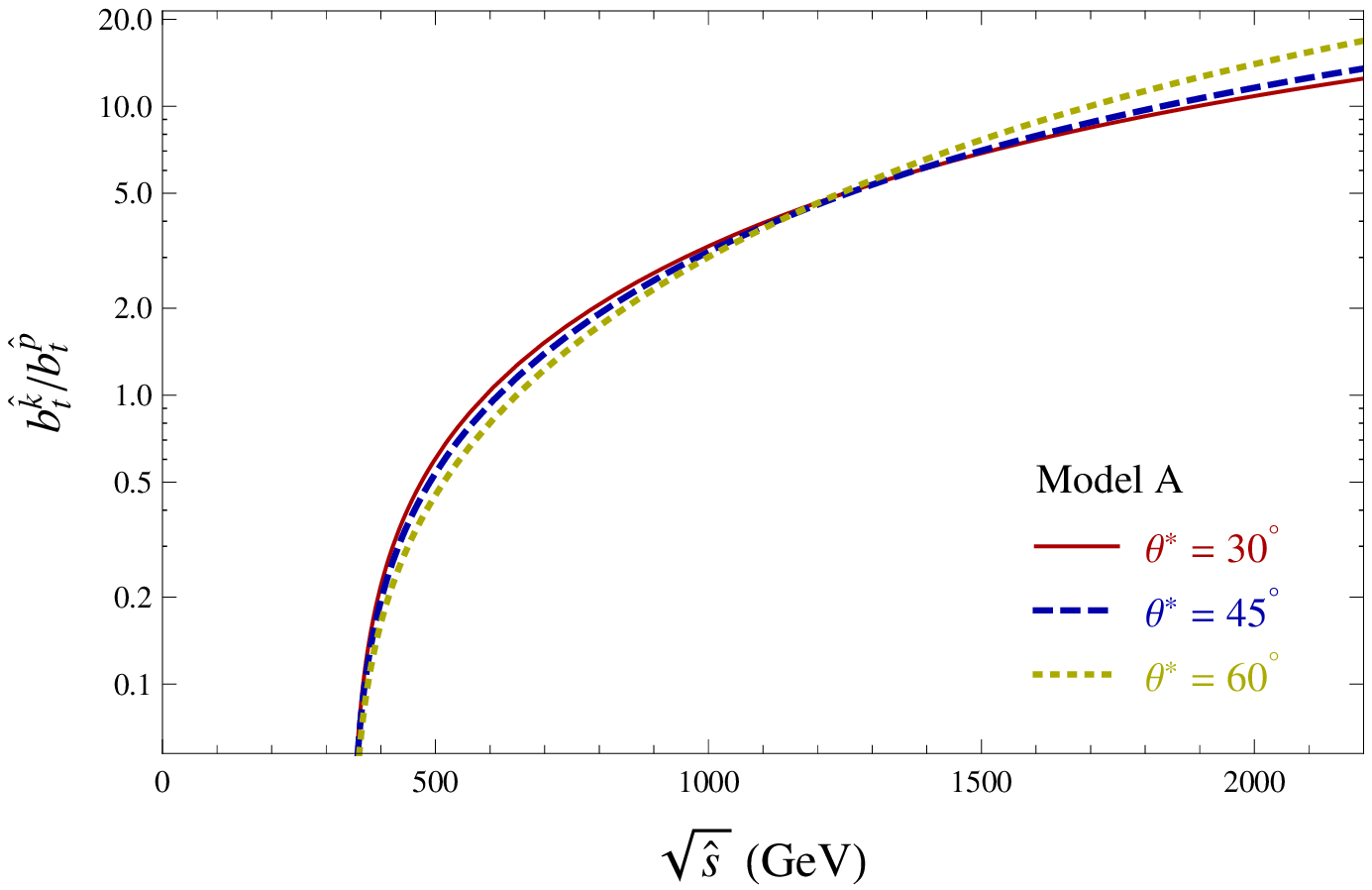}  
\includegraphics[width=0.46\textwidth]{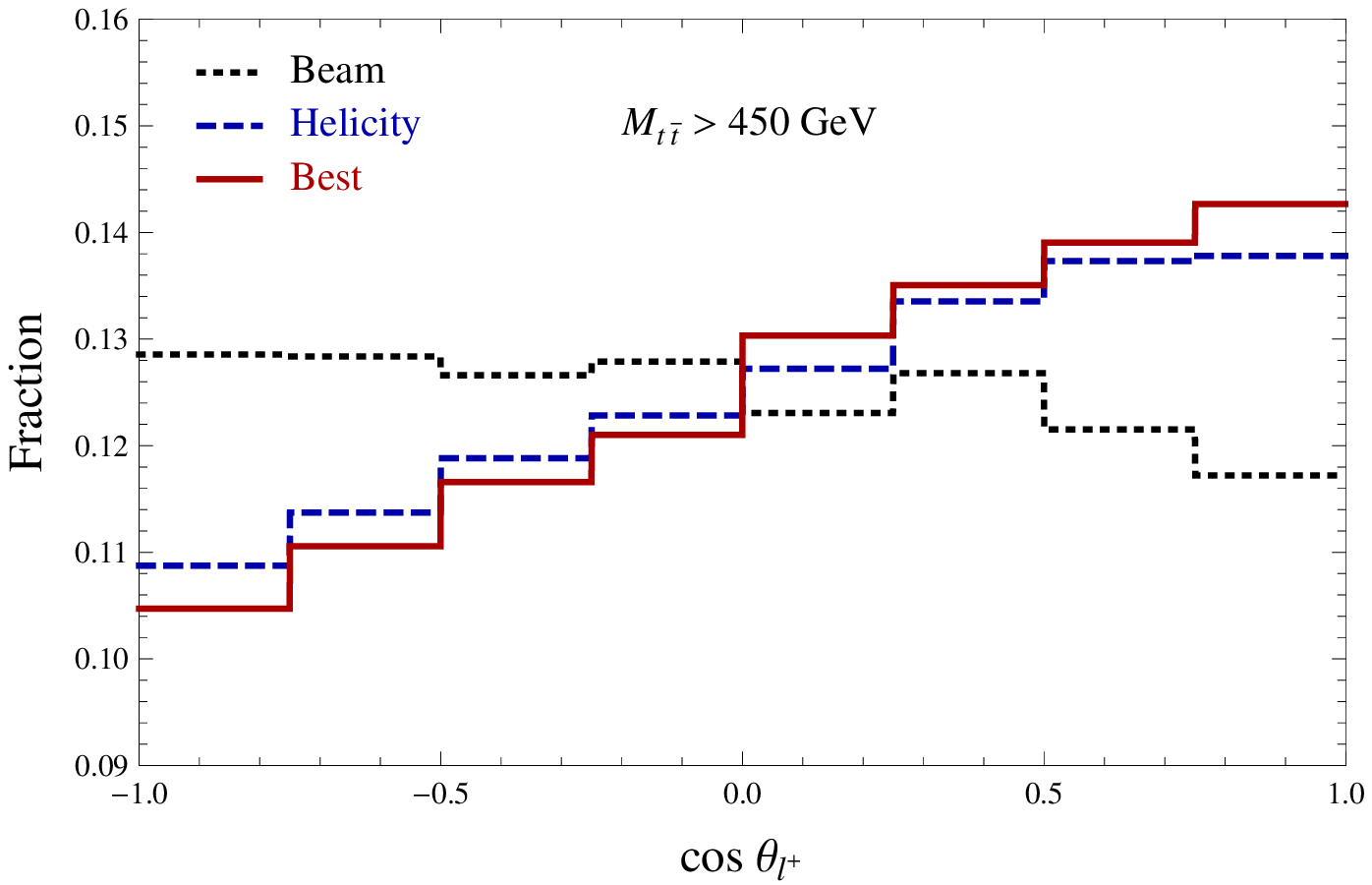} 
\caption{Left panel: the ratio $b_t^{\hat k}/b_t^{\hat p}$ of Model A, which determines the best top spin quantization axis, for different kinematics. Right panel: the normalized distributions of the angle between the lepton momentum in the top rest frame and different top spin quantization axes. All momenta are first boosted into the $t\bar t$ zero momentum frame and only the signal events, $t\bar t$ produced from light quarks, are included in this plot.}
\label{fig:PbkbpA}
\end{center}
\end{figure}
In the left panel of Fig.~\ref{fig:PbkbpA}, we show ratios of $b_t^{\hat k}$ over $b_t^{\hat p}$ as a function of $\hat s$ for different production angles. For $\hat s$ between 500 GeV to 1 TeV, $b_t^{\hat k}$ and $b_t^{\hat p}$ are comparable to each other, then the best spin quantization axis is neither the helicity basis nor the beam axis. So, using the best spin quantization axis may significantly increase the top polarization measurement and help improve the $A_{FB}$ measurement. In the right panel of Fig.~\ref{fig:PbkbpA}, we compare the sizes of the top quark polarization for different spin quantization axes. One can see that the top quark is indeed polarized in Model A. Since the background top quarks from $gg$ are not polarized, the corresponding $\cos{\theta_{\ell^+}}$  distribution is flat. Therefore one can use polarization effects to distinguish signals from backgrounds. From this plot one also sees that the ``best quantization axis" defined in Eq.~(\ref{eq:bestModelA}) gives us  a larger polarization effect than the other two axes: the helicity basis and the beam line basis. The differences could be even larger assuming we know exactly the $u$ or $d$ parton directions in each event. Unfortunately, the LHC is  a proton-proton collider and there is a $\sim20\%$ probability that we misidentify the initial parton moving directions.  

\begin{figure}[th!]
\begin{center}
\vspace*{3mm}
\includegraphics[width=0.46\textwidth]{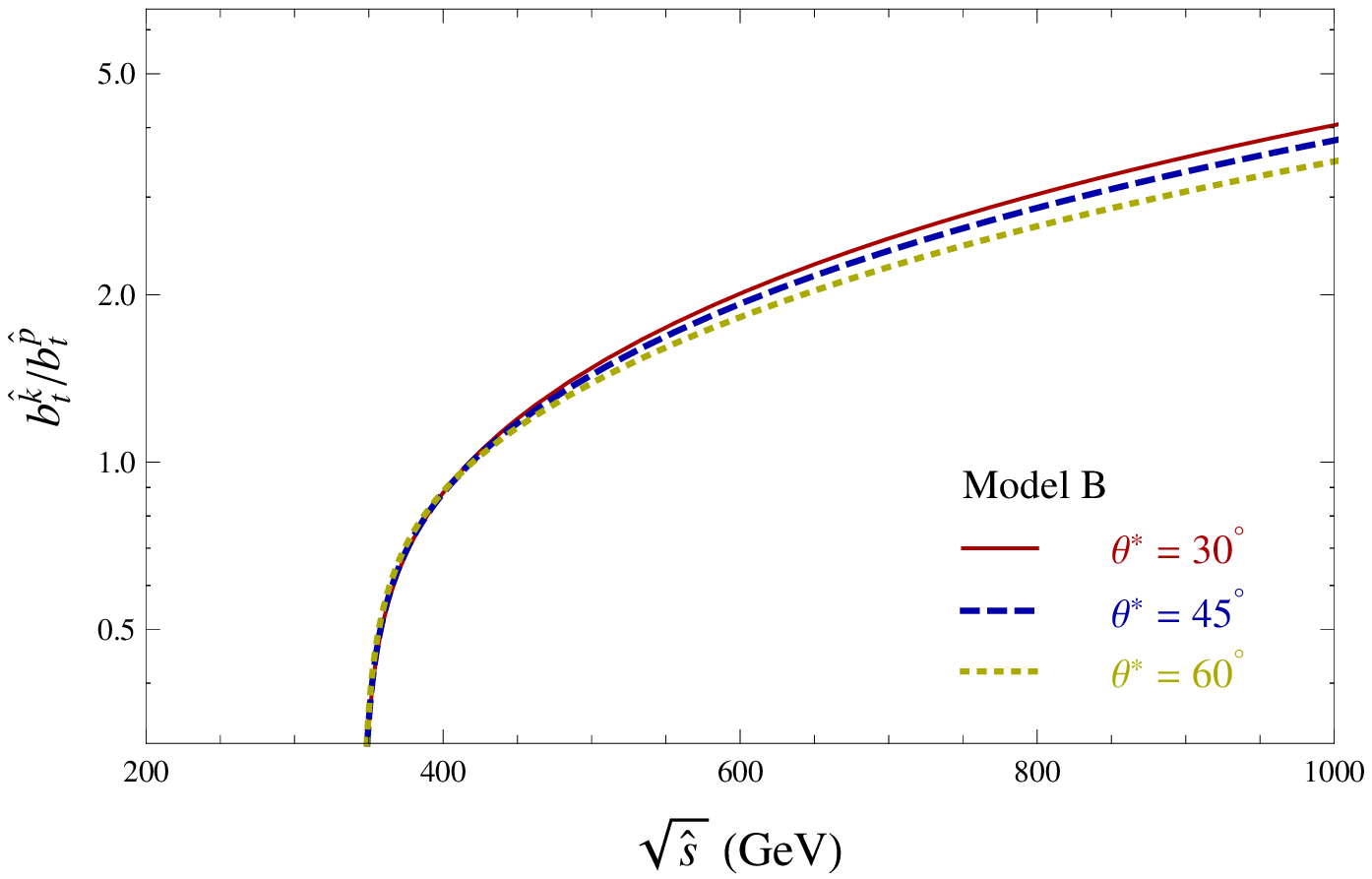}  
\includegraphics[width=0.46\textwidth]{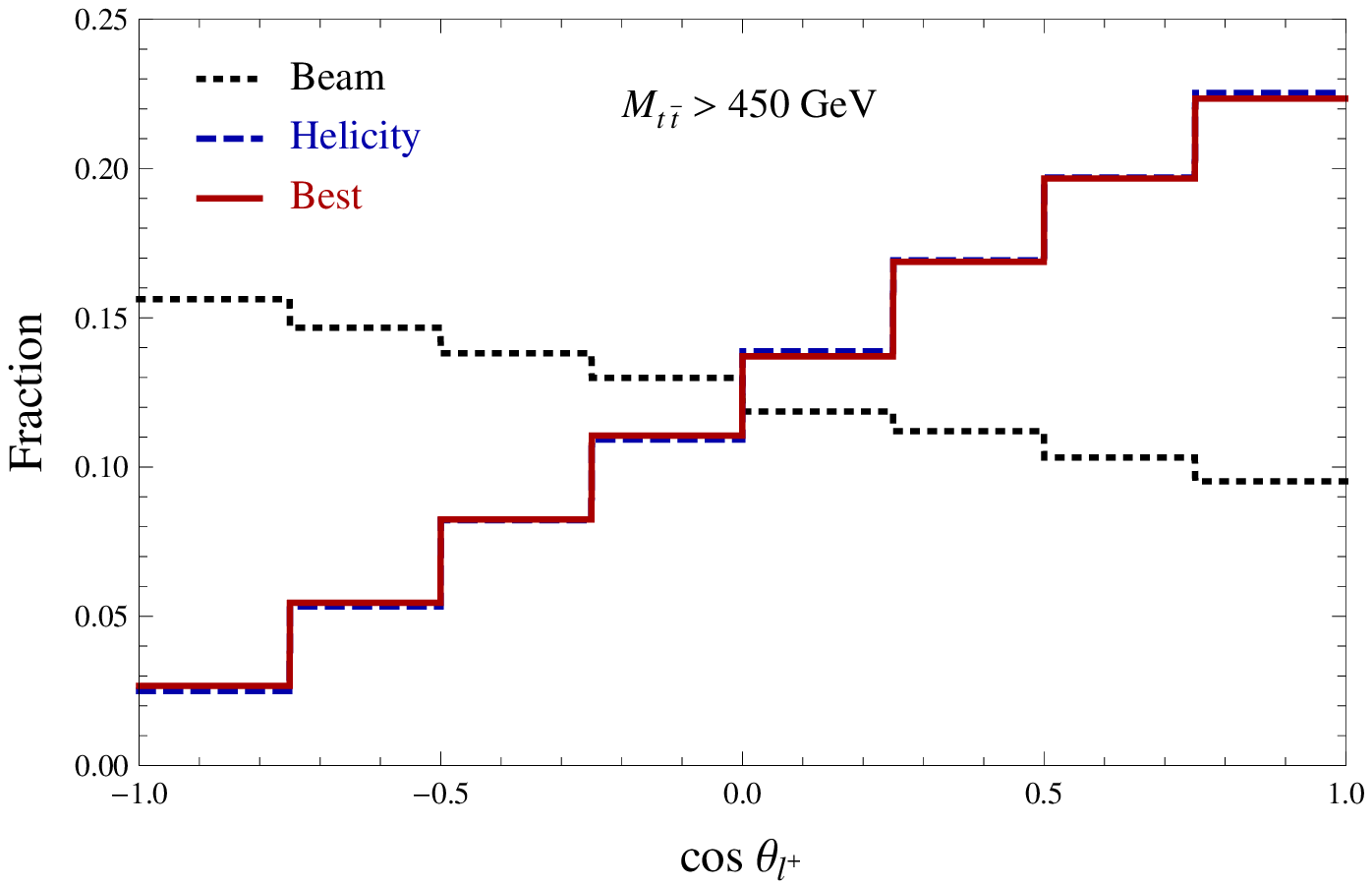} 
\caption{The same as Fig.~\ref{fig:PbkbpA}, but for Model B.}
\label{fig:PbkbpB}
\end{center}
\end{figure}

Let us turn to Model B. The $W^\prime$ boson in Model B only has right-handed couplings to quarks. A large polarization effect is anticipated in this model. Similar to the Model A case, we use $b_t^{\hat k}$ and $b_t^{\hat p}$ to define the best quantization basis for top quark polarization:
\beqa
b_t^{\hat k} &=& \frac{\beta  {g^2_R} s }{16 (2 m+1) {M^4_{W^\prime}}
   \left({M^2_{W^\prime}}-t\right)^2} \times 
  \left\{ 9 {g^2_R} m^4 s^3 (\beta  z-1) (2
   m-\beta  z+1) \right. \nonumber \\
   && \left.  -\,8 m^2 {M^2_{W^\prime}} s \left[9 \beta  {g^2_R} m s z+2
   t \left(2 m+(\beta  z-1)^2\right)\right] -32 {M^6_{W^\prime}} \left[2
   m+(\beta  z+1)^2\right] \right.   \nonumber \\
  && \left. +4 {M^4_{W^\prime}} \left[9
   {g^2_R} s (\beta  z+1) (2 m+\beta  z+1)   +8 m^3 s+4 m^2 s (\beta 
   z-1)^2+16 m t+8 t (\beta  z+1)^2\right] \right\} 
   \,, \nonumber \\
   b_t^{\hat p} &=&
\frac{{g^2_R} m s }{8 {M^4_{W^\prime}} \left({M^2_{W^\prime}}-t\right)^2} \times 
\left\{-9 {g^2_R} m^4 s^3 (\beta  z-1)+4 m^2 {M^2_{W^\prime}} s \left(9
   {g^2_R} s-4 \beta  t z+8 t\right) \right.  \nonumber \\
   && \left. -32 {M^6_{W^\prime}} (\beta 
   z+2)  +4
   {M^4_{W^\prime}} \left(9 {g^2_R} (\beta  s z+s)+4 m^2 s (\beta 
   z-2)+8 \beta  t z+16 t\right) 
   \right\}   \,.
\label{eq:bestModelB}
\eeqa
Here, $t = -\frac{1}{4} s (1+ \beta^2 - 2 \beta z)$, $z = \cos{\theta^*}$ and $m \equiv m_t/\sqrt{s}$. The right-handed gauge coupling $g_R$ is normalized with respect to the QCD coupling $g_s$. To derive those formulas, we have neglected the $W^\prime$ widths. The ratios $b_t^{\hat k}/b_t^{\hat p}$ as a function of the $\hat{s}$ for different production angles are shown in the left panel of Fig.~\ref{fig:PbkbpB}. The comparison of the top polarization for three different spin-quantization axis is shown in the right panel of Fig.~\ref{fig:PbkbpB}. One can see that there is not much difference between the ``best axis" and the helicity basis. The reason is that in this model most top pairs are produced with large center-of-mass energies, which can be seen from Fig.~\ref{fig:Pmttrapidity}. Because of the chiral coupling, in the massless limit the top quark has a definite helicity, {\it i.e.}, it is 100\% polarized in the helicity basis. Therefore, the large momentum of the top quark makes its mass  unimportant and the helicity basis close to the ``best axis".
The top quarks are largely polarized for Model B as can be seen from the right panel of Fig.~\ref{fig:PbkbpB}. Therefore, imposing a cut on $\cos{\theta_{\ell^+}}$ may improve the $A_{FB}$ measurement a lot.

%%%%%%%%%%%%%%%%%%%%%%%%%%%%%%%%
\section{Top and anti-top quark spin correlation}
\label{sec:spincorrelation}
Top quarks are not polarized in Model C, but the spins of the top and the anti-top are correlated. In this section, we show how one can obtain the correlation information and use it to distinguish between events from $gg$ and from $q\bar q$.

\begin{figure}[th!]
\begin{center}
\vspace*{3mm}
\includegraphics[width=0.46\textwidth]{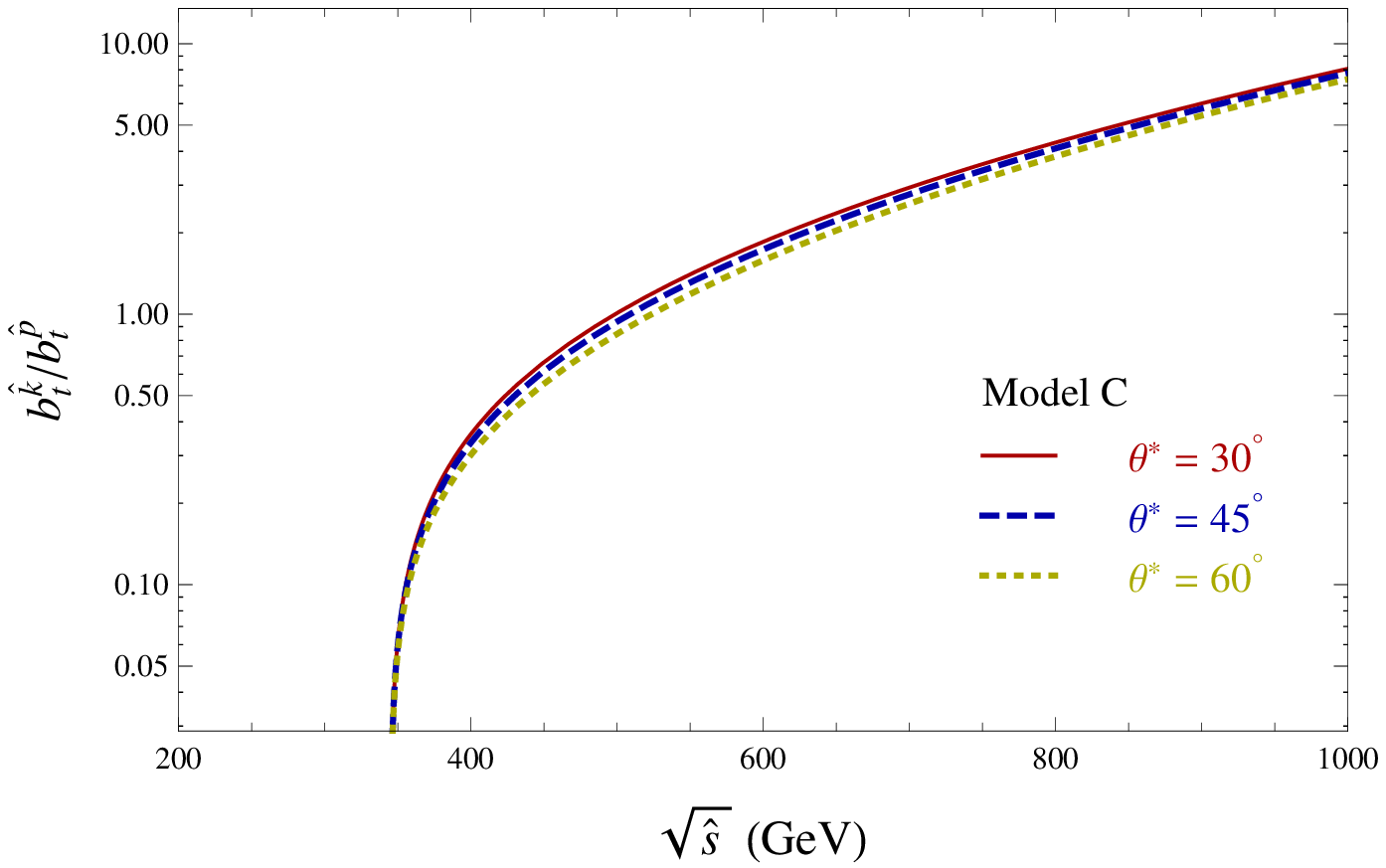}   \hspace{2mm}
\includegraphics[width=0.46\textwidth]{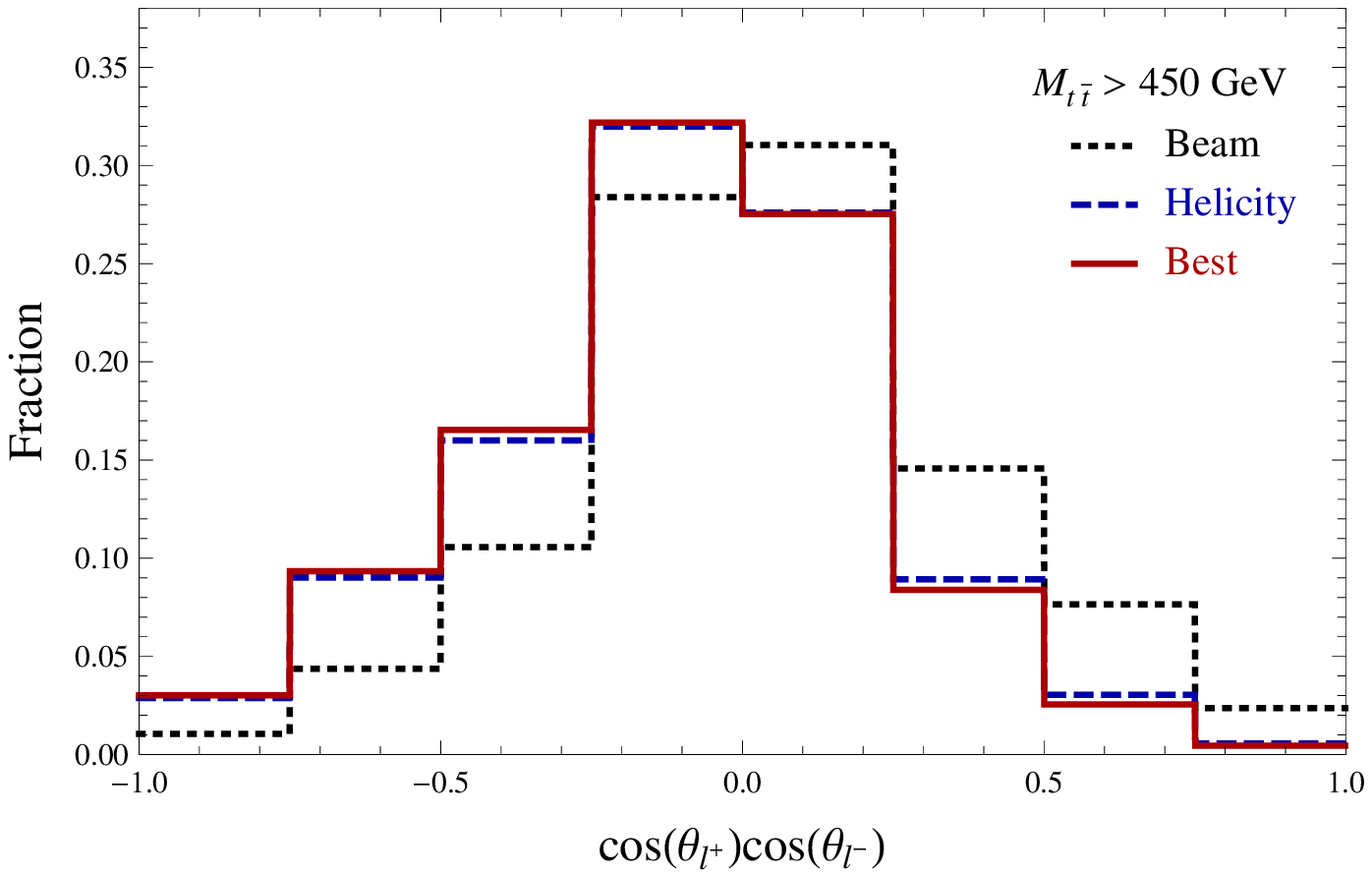} 
\caption{Left panel: the same as the left panel in Fig.~\ref{fig:PbkbpA}, but for Model C. Right panel: the normalized event distributions in $\cos{\theta_{\ell^+}}\cos{\theta_{\ell^-}}$ for different spin quantization axes.}
\label{fig:PbkbpC}
\end{center}
\end{figure}

The spin correlation of $t\,\bar{t}$ can be measured by studying the following double differential distributions
\beq
\frac{1}{N} \frac{d^2N}{d\cos{\theta_i} \,d\cos{\theta_j} } = \frac{1}{4} ( 1 - {\cal C}\, k_i k_j \,\cos{\theta_i} \,\cos{\theta_j} )\,,
\label{eq:two-dimension}
\eeq
The angles $\theta_{i}$ ($\theta_{j}$) is the angle between the quantization axis in Eq.~(\ref{eq:cutoffaxis}) and the daughter particle's momentum from top (anti-top) decay, measured in the top (anti-top) rest frame\footnote{One should first boost every momentum into the $t\bar{t}$ rest frame, and then boost the particles in the final state into the top/anti-top quark rest frame.}. The coefficients $k_i$, $k_j$ are constants determined by the particle species~\cite{Mahlon:1995zn}. Instead of fitting the distribution in terms of two variables $\cos{\theta_i}$ and $\cos{\theta_j}$, one can integrate Eq.~(\ref{eq:two-dimension}) to obtain the following one-dimensional distribution
\beq
\frac{1}{N} \frac{dN}{d[\cos{\theta_i} \cos{\theta_j}]} = \frac{1}{2}({\cal C}\, k_i k_j \,\cos{\theta_i} \cos{\theta_j} -1)
\log(|\cos{\theta_i} \cos{\theta_j}|)\,.
\label{eq:one-dimension}
\eeq
The parameter $\cal  C$ depends on the spin-quantization axis. The ``best axis" to maximize the spin-correlation for Model C is (for the detailed derivations, see Appendix~\ref{sec:appxContact})
\beqa
{\bf{e}}^q \propto {\bf \hat{p}} +  \left[ c_{\theta^*} (\gamma-1) - \frac{\xi\,\beta\gamma \hat{s}}{\Lambda^2}  \right]\,{\bf \hat{k}} \,, 
\label{eq:cutoffaxis}
\eeqa
In the left panel of Fig.~\ref{fig:PbkbpC}, we show the ratios of the ${\bf \hat{k}}$ component  over the ${\bf \hat{p}}$ component for different $\hat{s}$ and $\theta^*$.
In the right panel of Fig.~\ref{fig:PbkbpC}, we compare the spin-correlation effects by using different spin-quantization axes. One can indeed see that using the ``best axis" can increase the spin correlation compared to the other two axes. If there is no spin-correlation, the distribution should be symmetric for positive and negative values. Using the analytic formula in Eq.~(\ref{eq:one-dimension}), we fit the simulated distributions and found that ${\cal C} = 0.47, 0.80, 0.94$ for the beam line, helicity and ``best" axes, respectively. 

\begin{figure}[th!]
\begin{center}
\vspace*{3mm}
\includegraphics[width=0.46\textwidth]{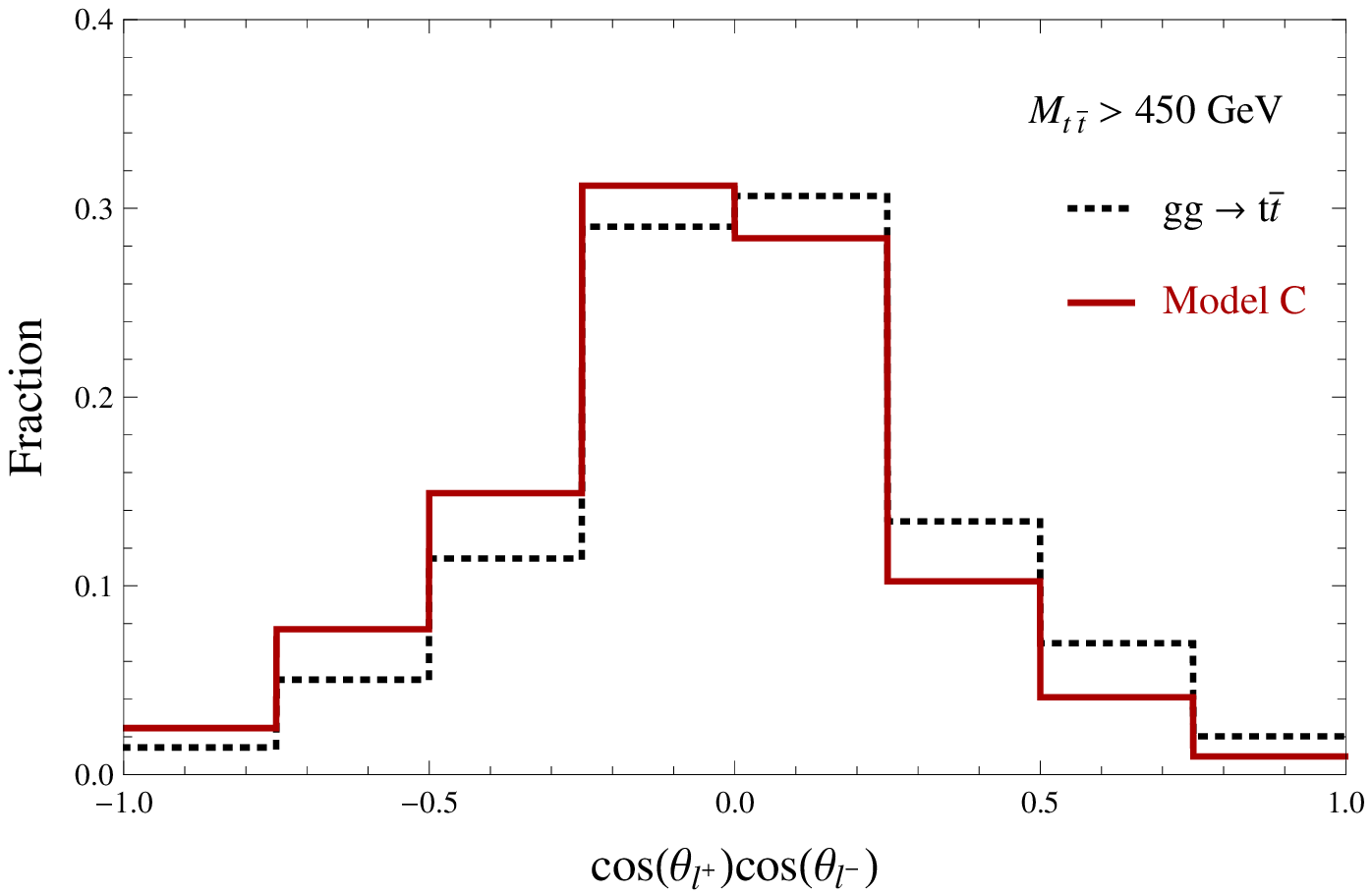}  \hspace{2mm}
\includegraphics[width=0.46\textwidth]{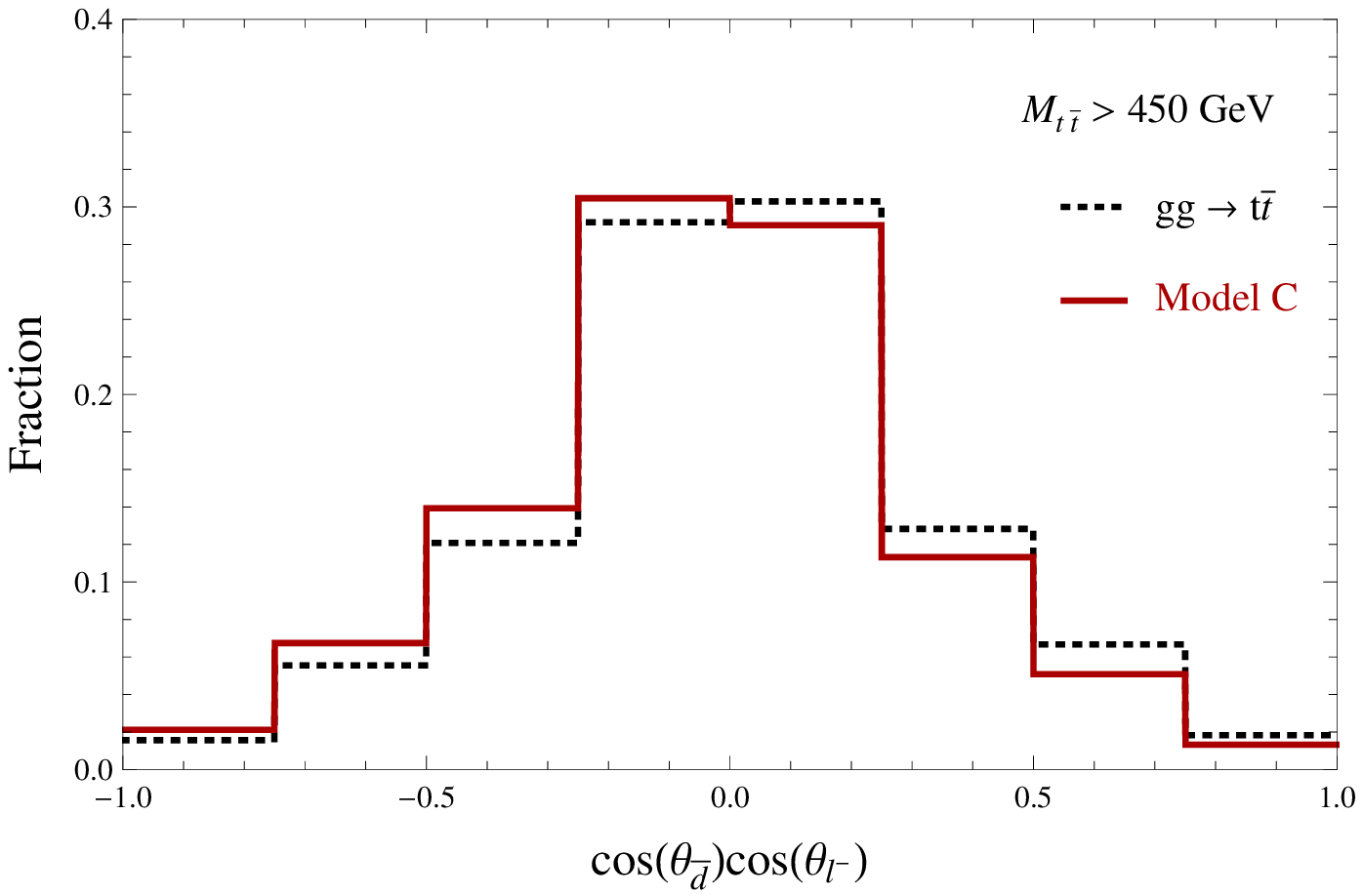}  
\caption{Left panel: the normalized event distributions of $\cos{\theta_{\ell^+}}\cos{\theta_{\ell^-}}$ for Model C and the $t\bar t$ events produced from $gg$ in the di-lepton channel by using the ``best axis". Right panel: the same as the left panel but for the semi-lepton channel.}
\label{fig:PCorrelSM}
\end{center}
\end{figure}

The background events from $gg\rightarrow t\bar t$ should also have spin-correlations for those three spin-quantization axes. In Fig.~\ref{fig:PCorrelSM}, we compare the signal .vs. background distributions using the ``best axis". 
One can see that their distributions are indeed different from each other, which can be used to improve the $A_{FB}$ measurement. For the di-lepton channel, we can use the two charged leptons to study the spin-correlation. After fitting the distributions, we have ${\cal C} = -0.13$ for $gg\rightarrow t\bar t$, which has an opposite sign to the signal, ${\cal C} = 0.94$. For the semi-leptonic channel, we can identify the jet closer to the $b$-quark in the $W^+$ gauge boson rest frame as the down-type quark (the probability is around 60\% from Ref.~\cite{Mahlon:1995zn}). After fitting to the distributions, we have ${\cal C} = -0.08$ for  $gg\rightarrow t\bar t$, which has an opposite sign to the signal, ${\cal C} = 0.41$. Since $gg\rightarrow t\bar t$ dominates the $t\bar t$ productions at the LHC, the spin correlation from the signal model will be diluted.making it difficult to measure the spin-correlation. However, eventually the LHC may accumulate enough data and make the spin-correlation measurement feasible, then one needs to find the optimal spin-quantization axis for the $gg\rightarrow t\bar t$ productions and the formulas obtained in this paper would be useful not only for the measurement but also for distinguishing between models.

%%%%%%%%%%%%%%%%%%%%%%%%%%%%%%%%
\section{Combined Improvement}
\label{sec:combined}
In this section, we combine the useful variables defined in the previous sections and consider the improvement on the $A_{FB}$ measurement.
% The improvement can be quantified in terms of the signal and background efficiencies, $\varepsilon_S\equiv n_S/n_S^0$ and $\varepsilon_B\equiv n_B/n_B^0$, where $n_S^0$ ($n_B^0$) is the number of signal (background) events we start with and $n_S$ ($n_B$) is the number of signal (background) events after cuts. Then we aim at maximizing $\varepsilon_S/\sqrt{\varepsilon_B}$ without the need to specify the original significance. 
Instead of using simple rectangular cuts, we adopt a likelihood discriminant method described as follows \cite{Aaltonen:2010jr}.

For a given variable $x^i$, we obtain from simulation the signal and background distributions as given in histograms $s^i$ and $b^i$. We normalize $s^i$ and $b^i$ such that they have the same binning and area. For a given event with the variable falling in the $j$'th bin, we define the probability of it being a signal event as 
\begin{equation}
p^i_s(x^i)=\frac{s_j^i}{s_j^i+b_j^i},
\end{equation}
where $s_j^i$ and $b_j^i$ are the numbers of events in the $j$'th bin for histograms $s^i$ and $b^i$ respectively. For multiple variables, the signal likelihood is defined as
\begin{equation}
\mathcal{L}_s= \frac{\Pi_i p^i_s}{\Pi_i p^i_s+\Pi_i(1-p^i_s)}.
\end{equation} 

Next we need to specify what variables to use when calculating the likelihood. The kinematic variables, $M_{t\bar t}$, $|y_t|$ and $|y_{t\bar t}|$ are useful for all models so we will always include them. Model B has large polarizations for the tops, so we include $\cos{\theta_{\ell}}$ and $\cos{\theta_{d}}$ using the best quantization axis for Model B, but not the spin correlation variables. The tops are not polarized in Model C, so we will use the spin correlation variable $\cos{\theta_{\ell}}\cos{\theta_{d}}$ based on the best quantization basis. For Model A, we use both the polarization variables and the correlation variable. For all variables, we group them to histograms with 20 bins. For $\cos\theta_\ell$, $\cos\theta_d$ and $\cos\theta_l\cos\theta_d$, the 20 bins have the same size from -1 to 1. For $M_{t\bar t}$, the first 19 bins have a size of $25~\gev$ ranging from $450~\gev$ to $925~\gev$ and the last bin contains all events with  $M_{t\bar t}>925~\gev$. For $|y_{t}|$ ($|y_{t\bar t}|$), the first 19 bins correspond to $(0, 1.9)$, evenly distributed, and the last bin contains all events with $|y_{t}| > 1.9$ ($|y_{t\bar t}|>1.9$).
\begin{figure}[th!]
\begin{center}
\begin{tabular}{ccc}
\includegraphics[width=0.33\textwidth]{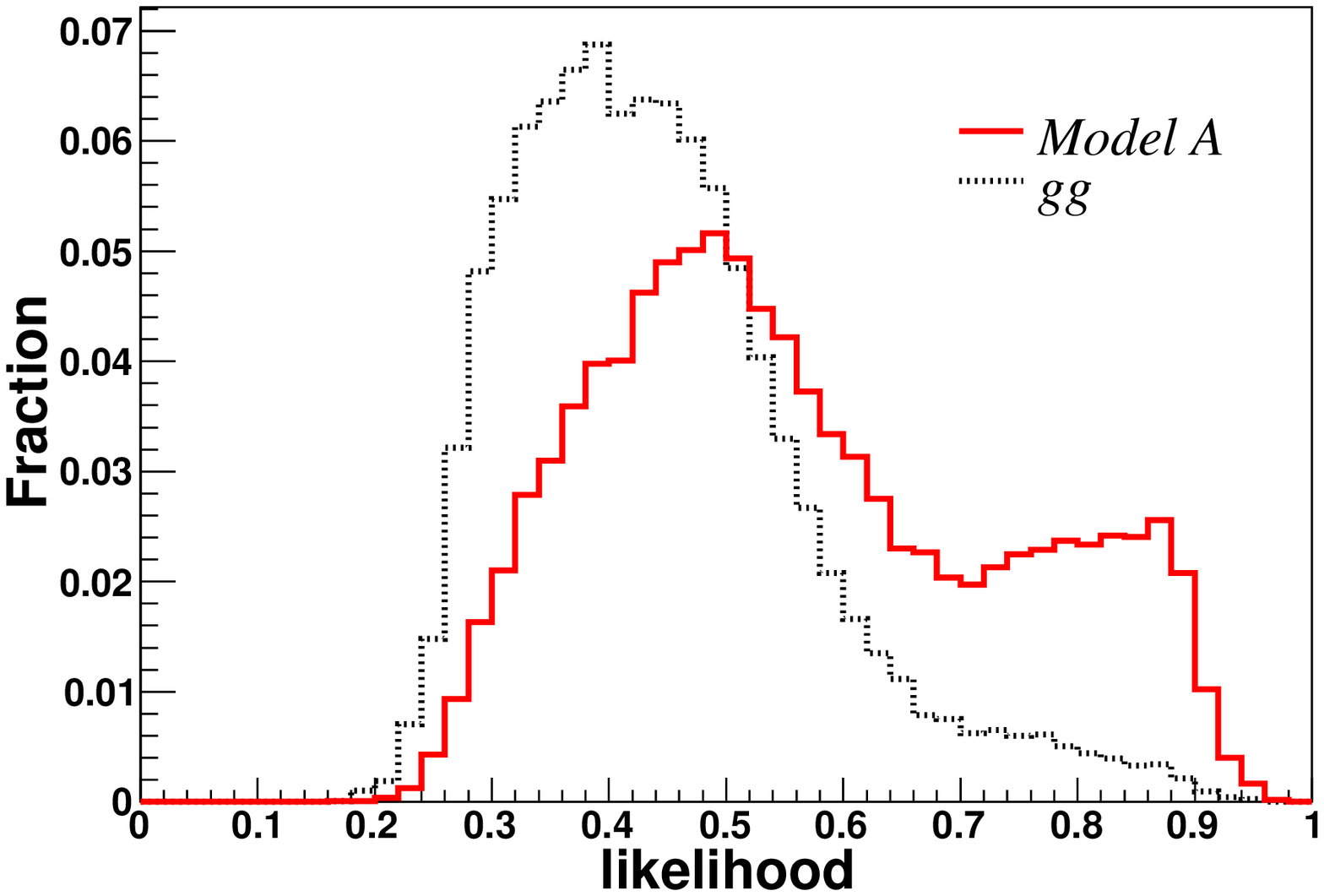}
&\includegraphics[width=0.33\textwidth]{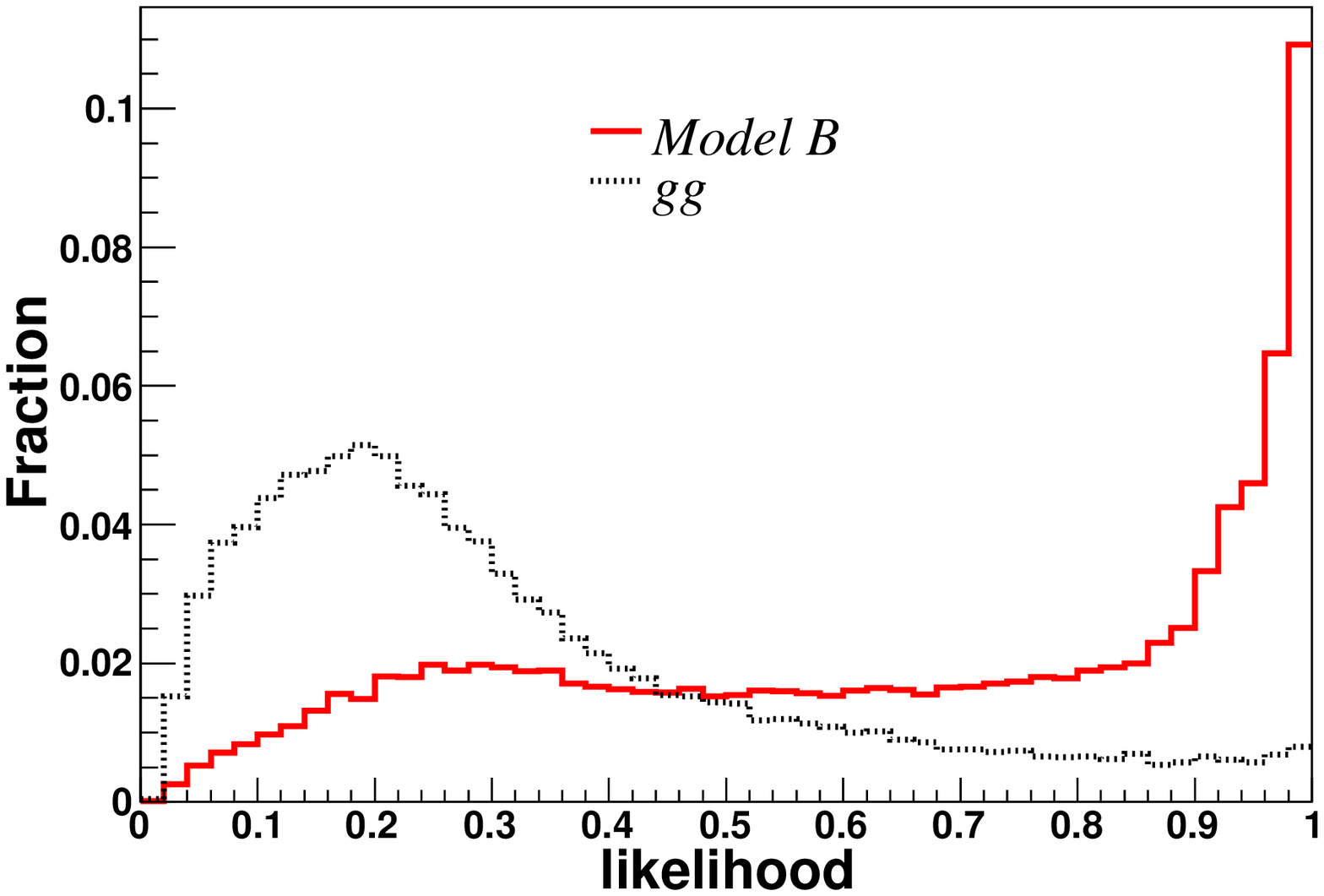}
&\includegraphics[width=0.33\textwidth]{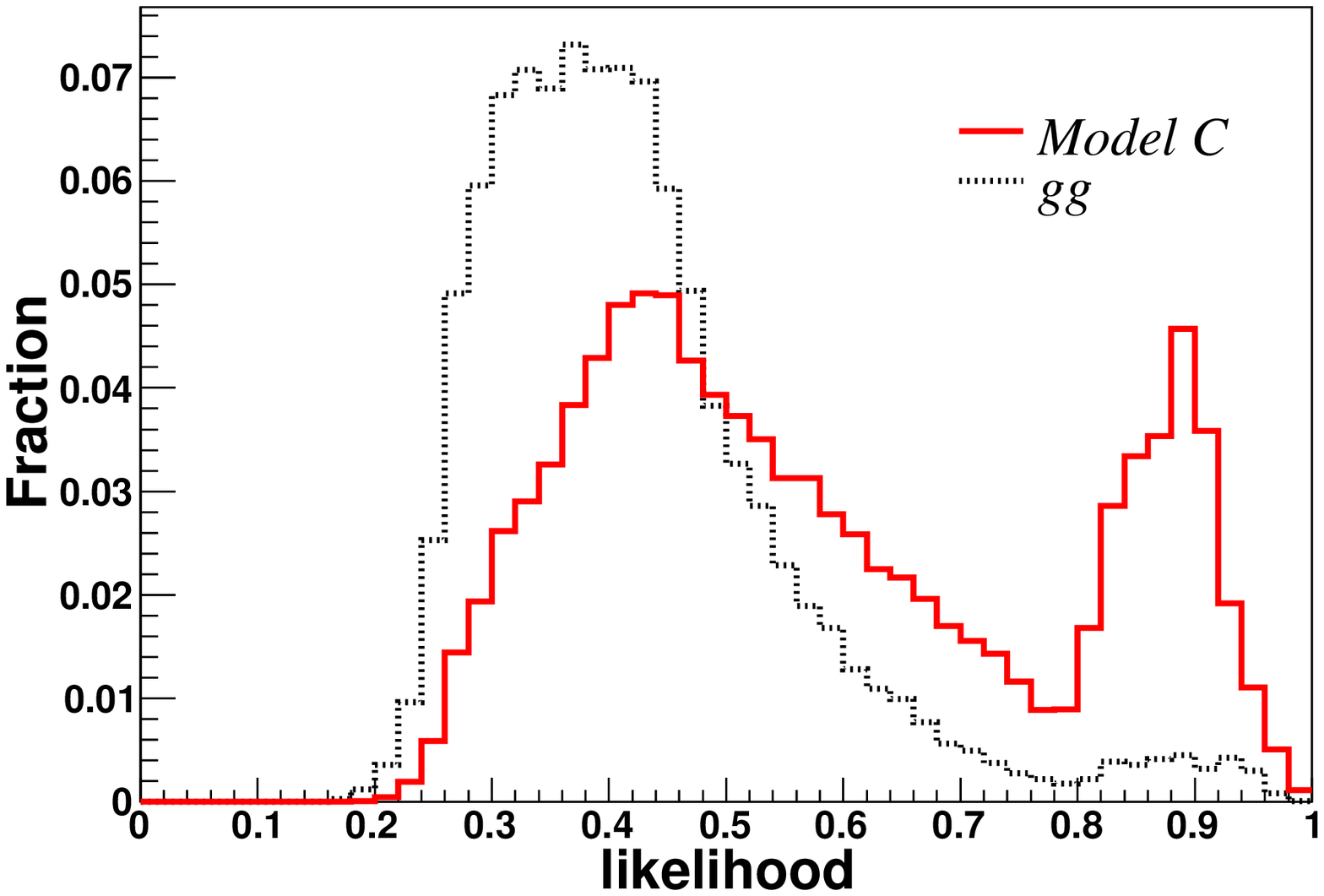}
\end{tabular}
\caption{Likelihood distributions for signal and background events.\label{fig:likelihood}}
\end{center}
\end{figure}

\begin{figure}
\begin{center}
\begin{tabular}{ccc}
\includegraphics[width=0.33\textwidth]{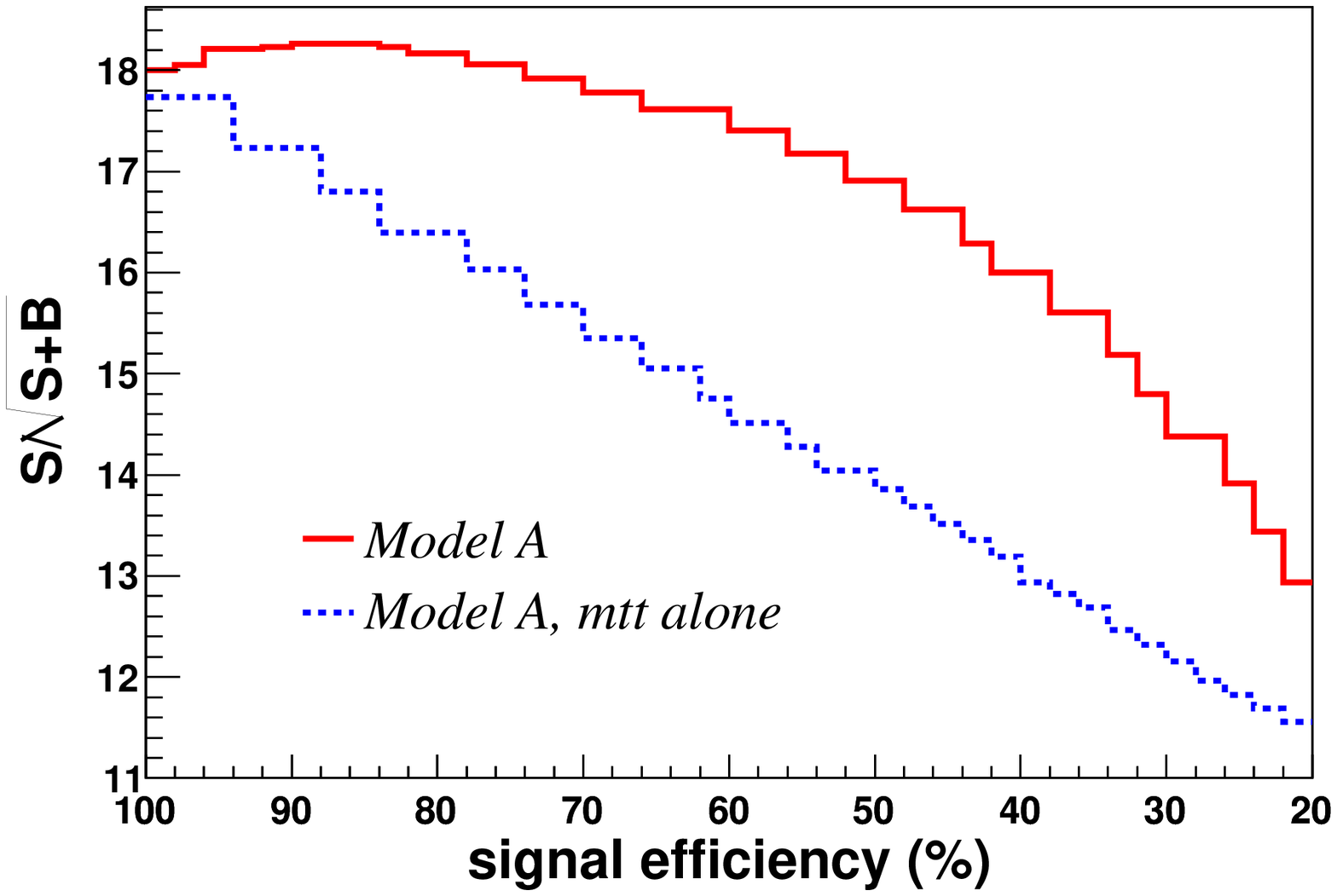}
&\includegraphics[width=0.33\textwidth]{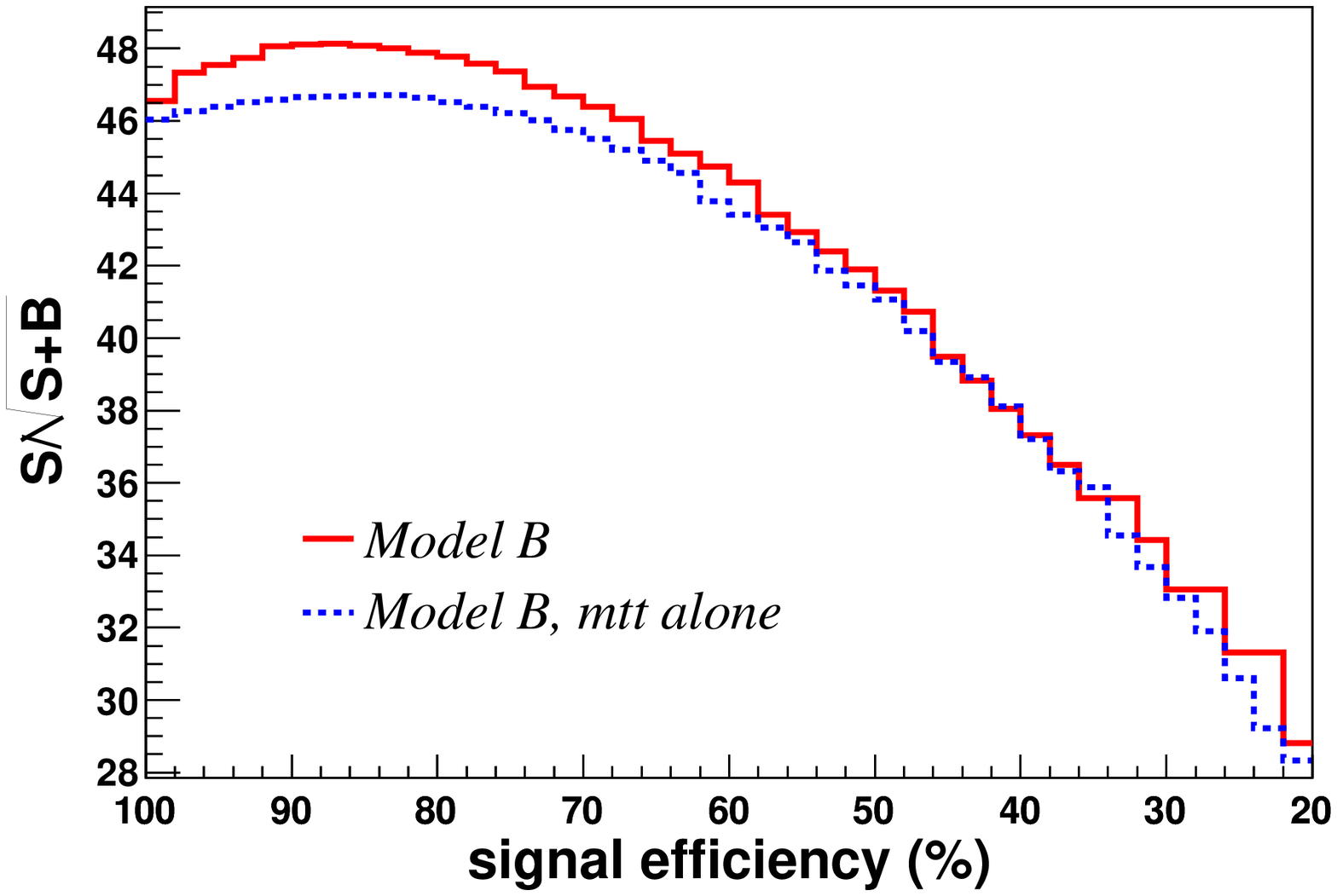}
&\includegraphics[width=0.33\textwidth]{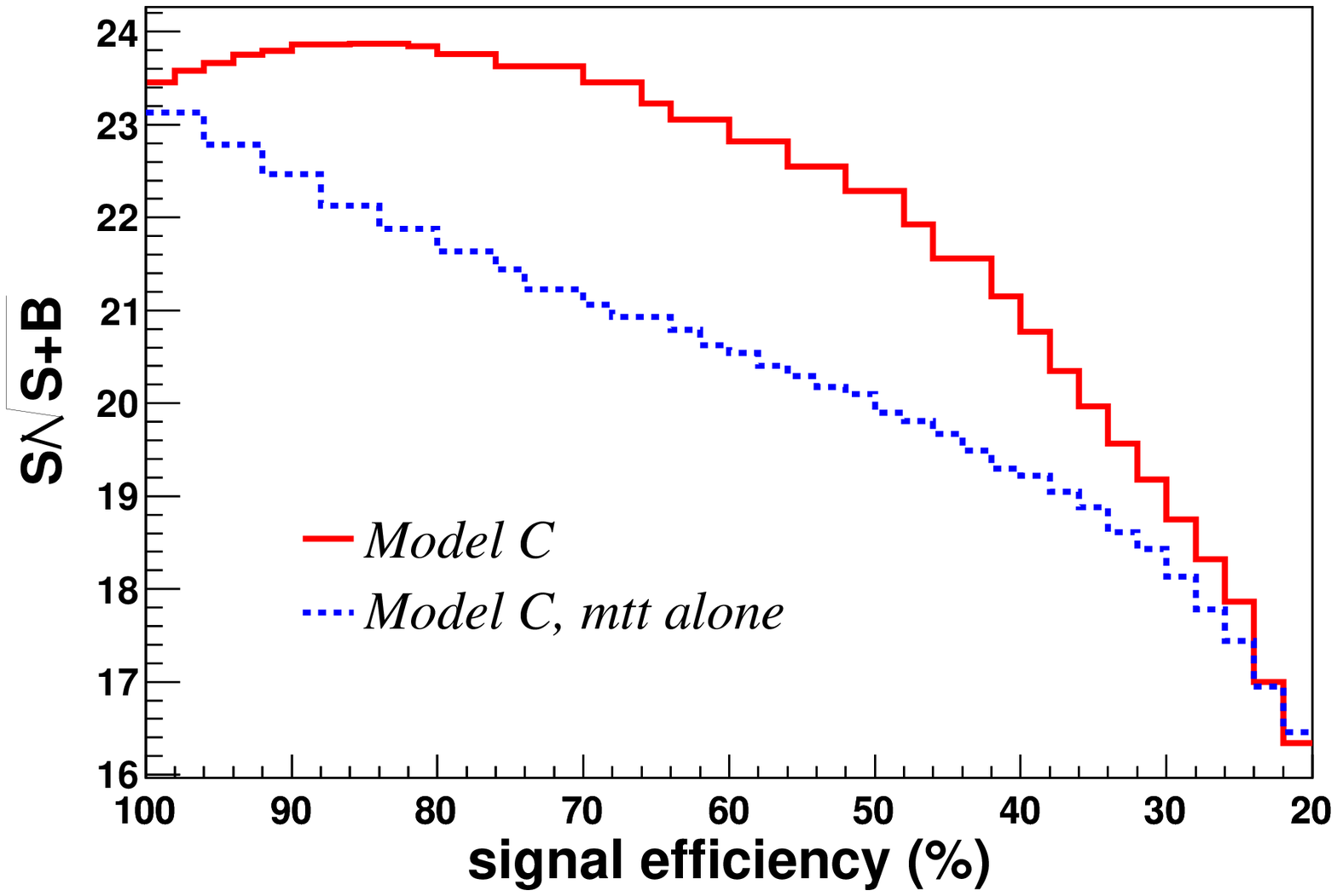}
\end{tabular}
\caption{Significance ($S/\sqrt{S+B}$) as a function of the signal efficiency. \label{fig:significance}}
\end{center}
\end{figure}

The signal and background likelihood distributions for the three models are shown in Fig.~\ref{fig:likelihood}. Given the distributions in Fig.~\ref{fig:likelihood}, we can choose a particular likelihood cut $\mathcal{L}_s^{\text{cut}}$  and keep events with $\mathcal{L}_s > \mathcal{L}_s^{\text{cut}}$. We then obtain the significance, $S/\sqrt{S+B}$, as a function of $\mathcal{L}_s^{\text{cut}}$. Since the signal efficiency is a monotonous function of the likelihood cut, we can change the variable and draw the the significance as a function of the signal efficiency instead, which is shown in Fig.~\ref{fig:significance}.  Note that the peaks at high likelihood values arise from the high $M_{t\bar t}$ tail, where the events are much more likely signal events.  The most efficient variable to distinguish signal and background is also $M_{t\bar t}$.  Therefore, it is illuminating to examine the significance improvement by cutting on $M_{t\bar t}$ alone and to compare it to the improvement by including all variables, which is also shown in Fig.~\ref{fig:significance}.
Another characteristic of the improvement is the signal-background ratio ($S/B$). Obviously, for the same significance, we would like $S/B$ as large as possible. This is due to two reasons. First, for larger $S/B$, the results will be less sensitive to the background systematic uncertainties. Second, since the background in our case has no asymmetry, for larger $S/B$, the measured central value of asymmetry will also be larger and deviate more from a flat distribution.  
\begin{figure}[th!]
\begin{center}
\begin{tabular}{ccc}
\includegraphics[width=0.33\textwidth]{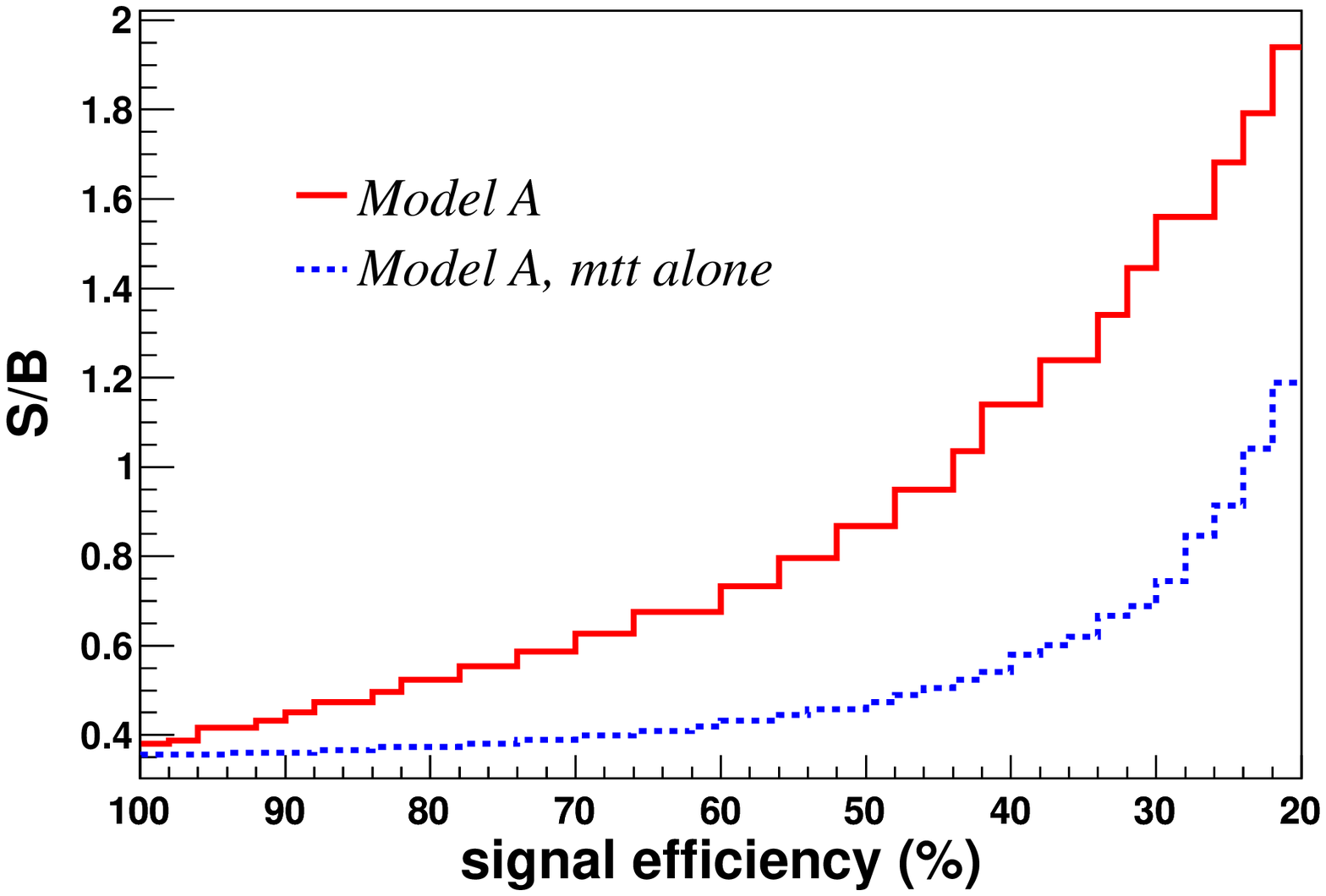}
&\includegraphics[width=0.33\textwidth]{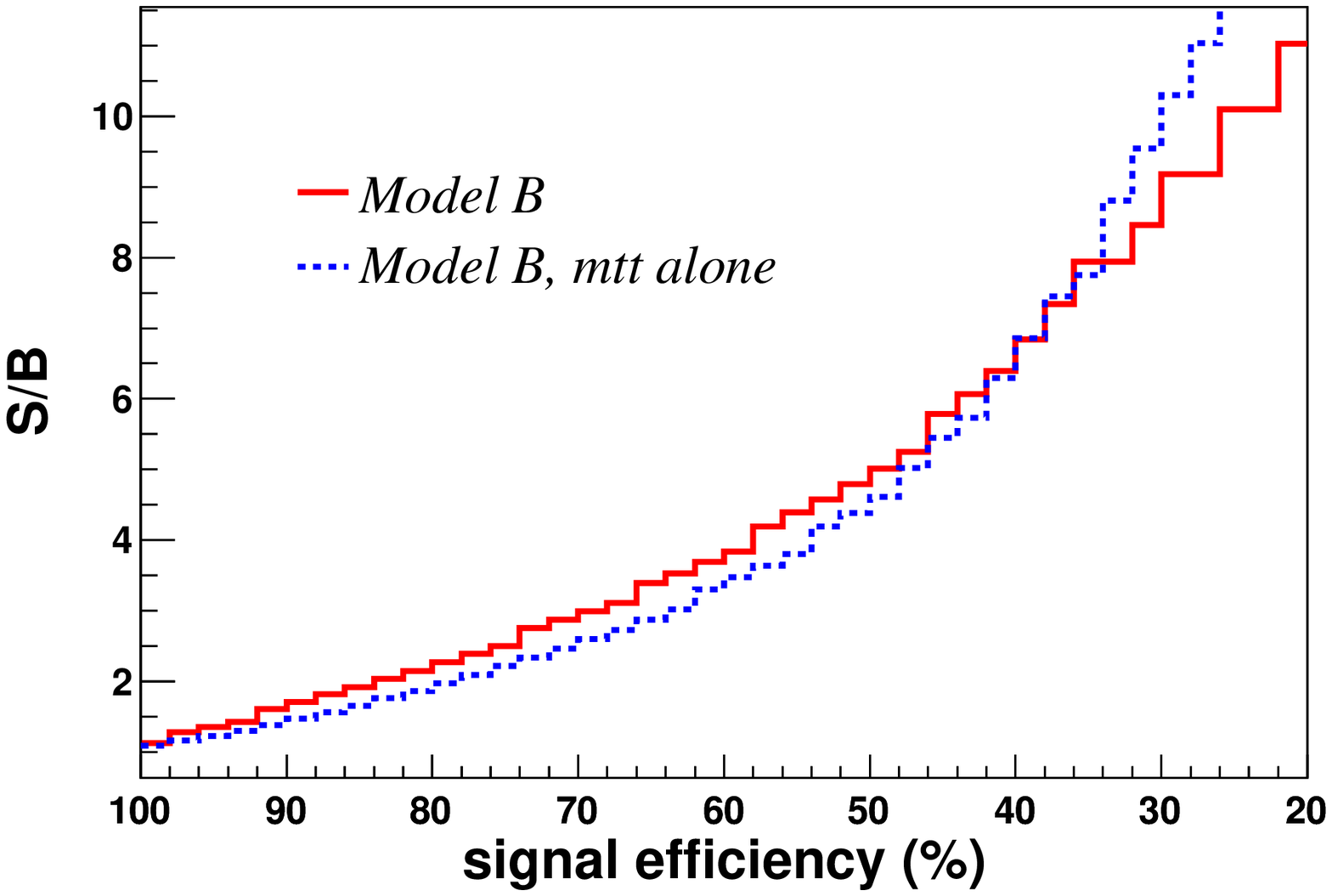}
&\includegraphics[width=0.33\textwidth]{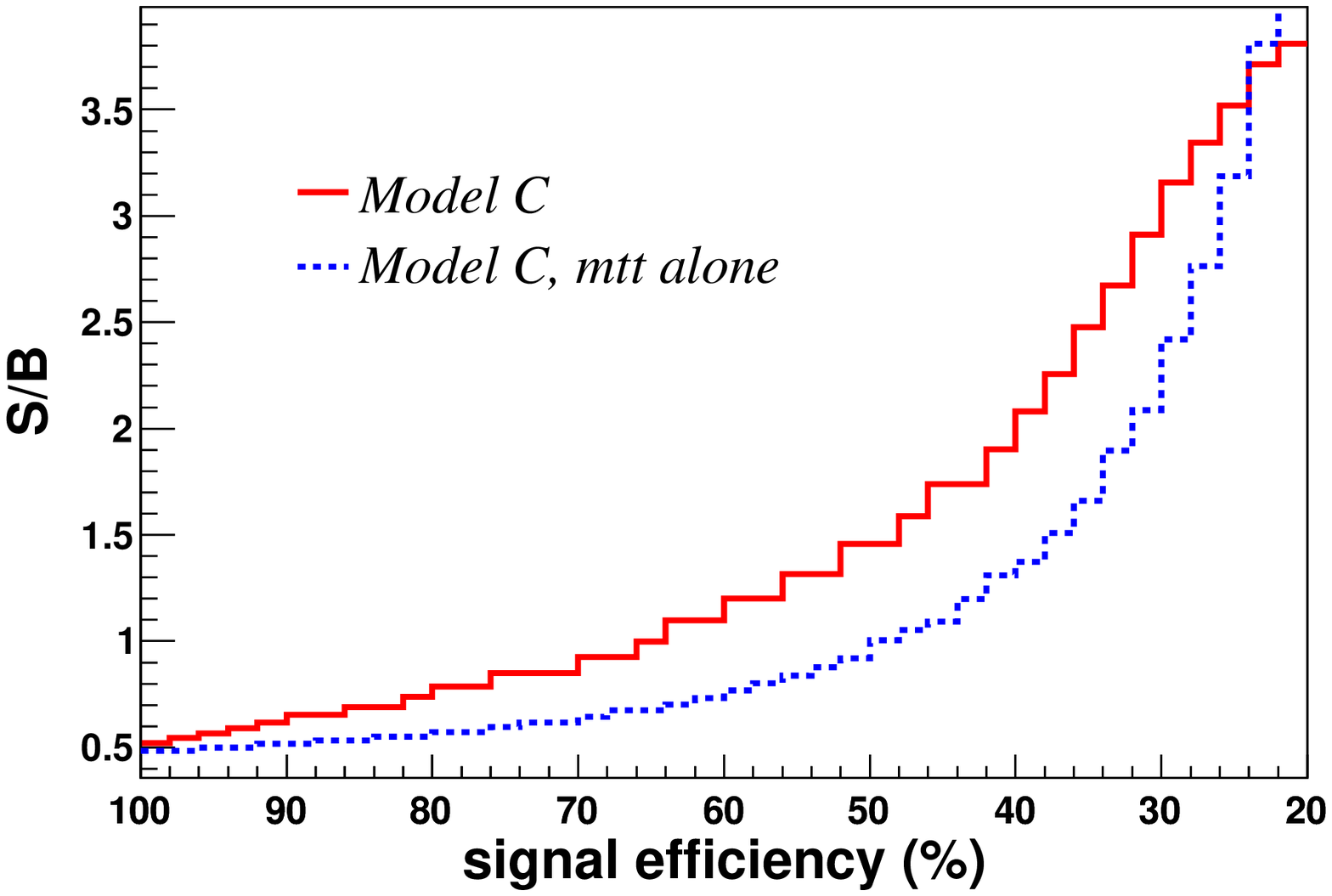}
%\\(a) &(b)
\end{tabular}
\caption{Signal-background ratio as a function of the signal efficiency. \label{fig:sb}}
\end{center}
\end{figure} 
\begin{figure}
\begin{center}
\begin{tabular}{ccc}
\includegraphics[width=0.33\textwidth]{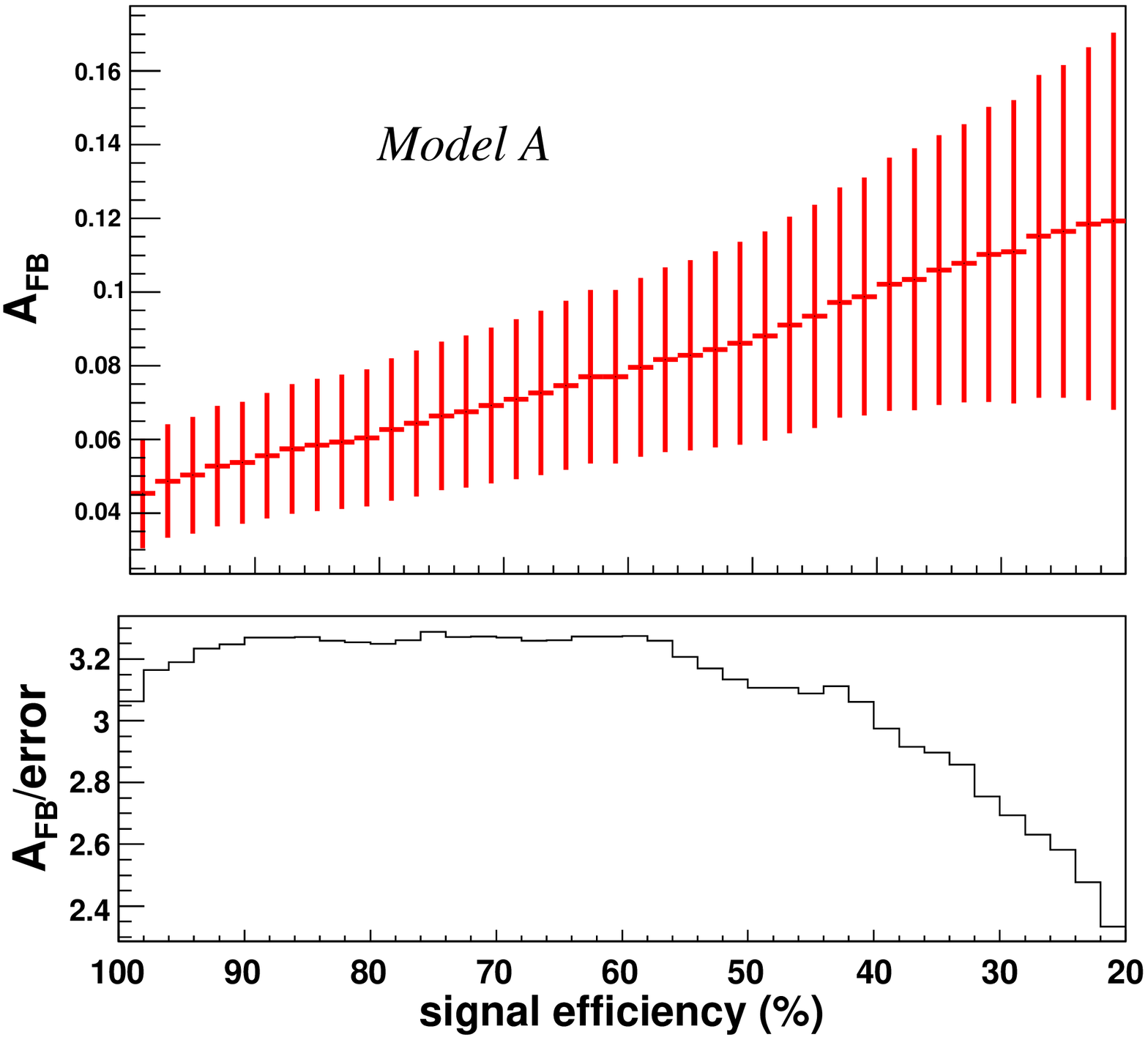}
&\includegraphics[width=0.33\textwidth]{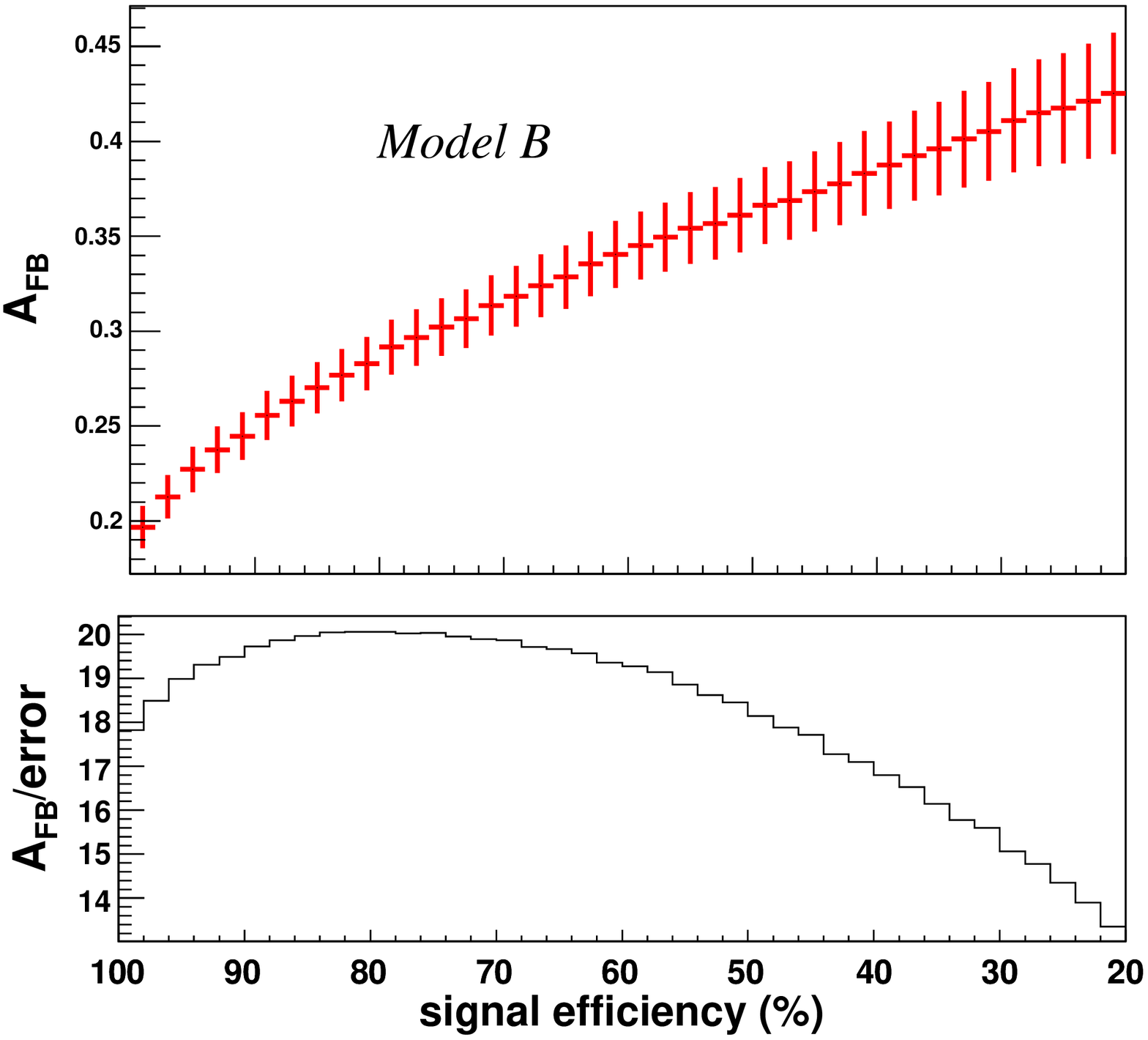}
&\includegraphics[width=0.33\textwidth]{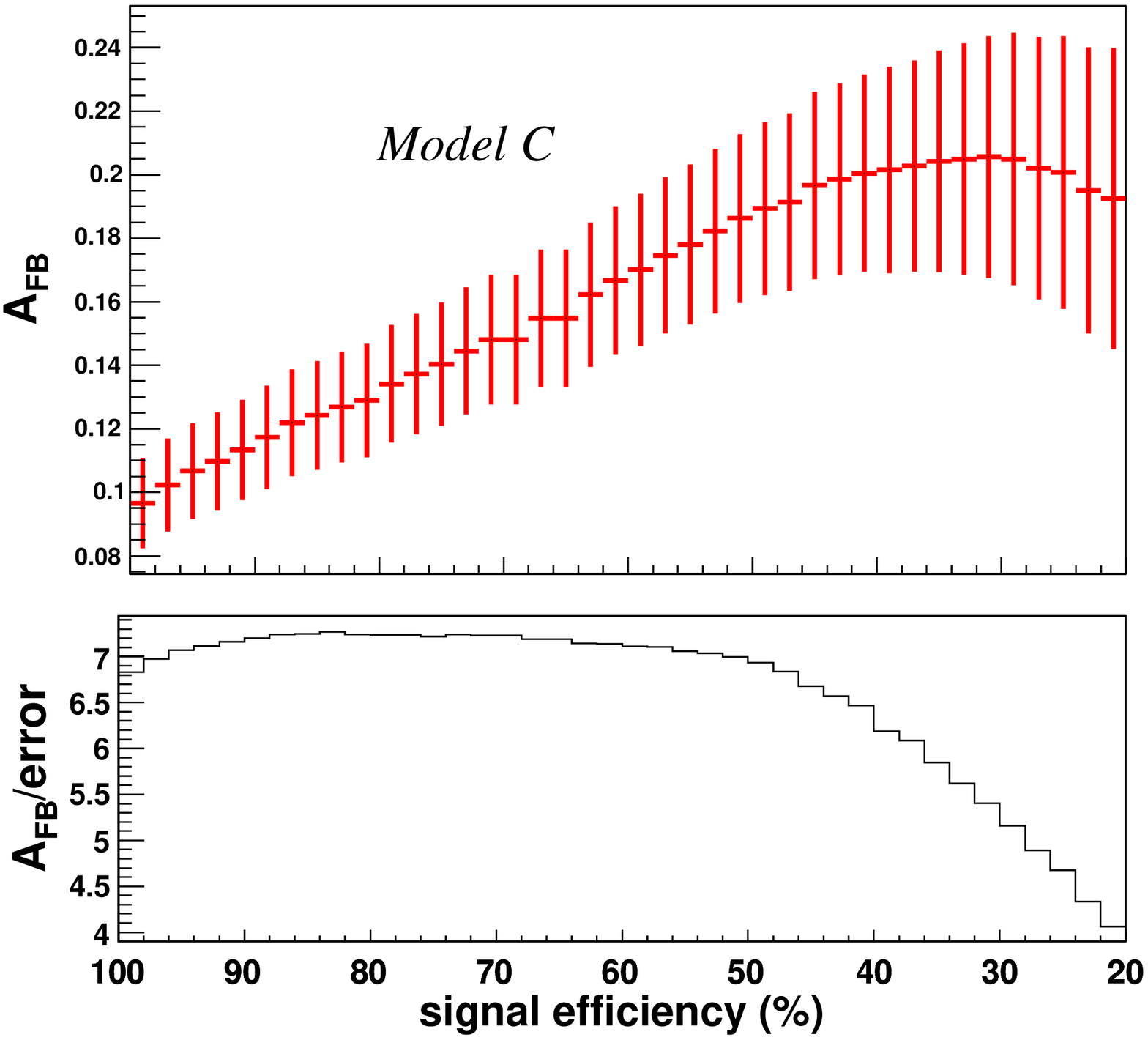}
\end{tabular}
\caption{Asymmetry central values and errors as a function of signal efficiency. \label{fig:asymmetry}}
\end{center}
\end{figure} 
From Fig.~\ref{fig:significance} and Fig.~\ref{fig:sb}, we see that we can increase $S/B$ and at the same time obtain moderate improvement in $S/\sqrt{S+B}$. This will help with the discovery of the signal events by making it less sensitive to systematic uncertainties. 

\begin{figure}
\begin{center}
\begin{tabular}{c}
\includegraphics[width=0.4\textwidth]{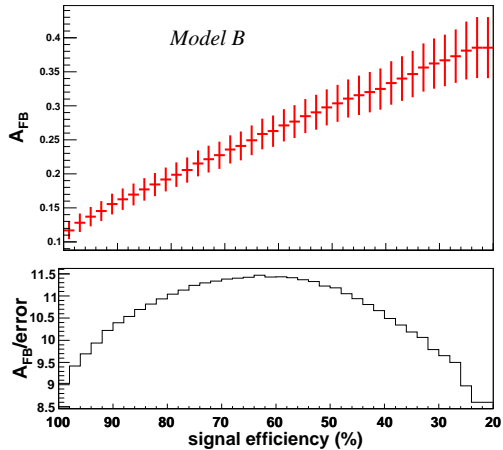}
\end{tabular}
\caption{Asymmetry central values and errors as a function of signal efficiency for Model B with a smaller coupling, $g_R=1.5$. \label{fig:asymmetry_b2}}
\end{center}
\end{figure}

On the other hand, we should make sure that the cuts we use do not significantly reduce the asymmetry. Therefore, we need to calculate the asymmetry using events which pass the likelihood cut and examine directly whether we have improved the asymmetry measurement or not. Moreover, the likelihood cut that maximizes $S/\sqrt{S+B}$ in general is not the cut that maximizes the asymmetry. Therefore we scan the likelihood cut and find the cut that maximizes $\afb/\sigma_{\afb}$, where $\sigma_{\afb}$ is the error for the asymmetry measurement given by $1/\sqrt{N}$ with $N$ the total number of events after cuts and after taking a 10\% acceptance into account. The asymmetries and errors, as well as their ratios as a function of signal efficiency are shown in Fig.~\ref{fig:asymmetry}. Then we find the best likelihood cuts for the three models are 0.49, 0.33 and 0.36 respectively, corresponding to signal efficiencies of 0.60, 0.80 and 0.86, and background efficiencies of 0.30, 0.37 and 0.63. The resulting asymmetries are given by
\beqa
\mbox{Model A:}&& \quad A_{FB}(\mathcal{L}_S > 0.49) = 0.078\pm0.024\,, \nonumber \\
\mbox{Model B:}&& \quad A_{FB}(\mathcal{L}_S > 0.33) = 0.289\pm0.014\,, \nonumber \\
\mbox{Model C:}&& \quad A_{FB}(\mathcal{L}_S > 0.36) = 0.121\pm0.017\,. 
\eeqa
Comparing with the numbers in Eq.~(\ref{nocut}), we see that we have achieved larger central values for the $\afb$ measurements with improved $\afb/\sigma_{\afb}$. Note that the improvement in $\afb/\sigma_{\afb}$ is not significant for all three models. This is due to different reasons: for Model A and Model C, the likelihood distributions for the signal and the background are not dramatically different; for Model B, although the distinction between the signal and the background is large, the signal cross section for $m_{t\bar t}$ is so large that it is not essential to reduce the number of background events. As mentioned previously, given the recent LHC results on $M_{t\bar t}$ distribution measurement, Model B is no longer viable unless the $W^\prime-t-d$ coupling, and therefore the  signal cross section are smaller. In that case, our method will be more useful. As an illustration, we consider Model B with the same $W^\prime$ mass but a coupling $g_R=1.5$ and repeat our optimization procedure. The signal cross section for $M_{t\bar t}>450~\gev$ is reduced to 21 pb from 38 pb of the original model. The resulting asymmetry as a function of signal efficiency is given in Fig.~\ref{fig:asymmetry_b2}. We see that $\afb/\sigma_{\afb}$ is improved by about 30\% for the best cut, with the central $\afb$ value more than doubled. 

%%%%%%%%%%%%%%%%%%%%%%%%%%%%%%%%
\section{Discussion and conclusions}
\label{sec:conclusion}
%%%%%%%%%%%%%%%%%%%%%%%%%%%%%%%%

For all new physics explanations of $A_{FB}$, the differential cross sections in $M_{t\bar t}$  have been predicted to be different from the SM. Measuring this distribution would be the first hint of new physics behind $A_{FB}$. However, the measurement of the $t\bar t$ production cross section as well as $d\sigma / d M_{t\bar t}$ distributions are suffering from a large systematic errors related to the jet energy scaling and the luminosity uncertainty~\cite{Atlasttbar}. If the resonance in the $s$-channel is very broad or the new particle in the $t$-channel does not contribute to the $t \bar t$ productions significantly, performing a precision measurement like measuring $A_{FB}$ could be the unique way to unravel the new physics behind top quarks. 

Treating the $t\bar t$ production from gluons as backgrounds and those from light quarks as signals, we have found that the central values of $A_{FB}$ can be increased by a factor as large as 2 for all three models when we impose stringent cuts on both signal and background events. The real improvement on the significance of measurement, on the contrary, can only be increased by 10\% to 30\%. The simple reason is that the optimized cuts from our likelihood analysis decrease the signal efficiency as quickly as increase the central values $A_{FB}$. We believe that this result is true not only for the measurement of  $A_{FB}$ but also for the charge asymmetry measurement because of a strong correlation between $A_{FB}$ and the top quark charge asymmetry. From Fig.~\ref{fig:sb}, one can see that the ratio of top quarks produced from light quarks over from gluons can be increased by 100\% with a moderate cost of signal efficiency. This improvement of $S/B$ can eventually help the $A_{FB}$ measurement especially when the systematical errors are large. 

In all of our analysis above, we have neglected the effects on our results from the top reconstruction procedure as well as various experimental cuts on the final state particles from the top decays. Especially, the top polarization and top spin-correlations will be affected by the rapidity cuts on leptons and jets. The actual difference between the signal and background distributions in Fig.~\ref{fig:PCorrelSM} will be reduced. To find more accurate experimental improvement in the $A_{FB}$  measurement using the variables in this paper, one also needs to consider other effects such as hadronization, initial state radiation, final state radiation and the detector effects, which are not included in this paper.  

Another concentration of this paper is to find the best quantization axis for studying the top quark polarization and top-antitop spin-correlation. For the new physics models considered in this article, we have obtained the axis giving us the maximal top polarization, as well as the axis yielding maximal top-antitop spin correlation. If new particle contributions to top quark pair production do exist, one can scan different spin quantization axes to find the best one that maximizes the polarization or spin-correlation effects. Then, we can use the analytical formulas for the best quantization axis to fix a relation among the model parameters. From this point of view, finding those axes are not only helpful for improving $A_{FB}$ measurement but also useful for disentangling different new physics models. 

In conclusion, we have explored various kinematic distributions for improving the $A_{FB}$ measurement at the LHC. We have adopted a multivariate likelihood discriminant to obtain the potential improvement by including the invariant mass of the top pair, the rapidity of $t\bar t$ system in the lab frame, the rapidity of top quarks in the $t\bar t$ rest frame, as well as top polarization and top-antitop  spin correlation. Treating top pairs produced from gluons as the background and those from light quarks as the signal, we have found that the $A_{FB}$ measurement can be improved by $10\% - 30\%$ depending on the underlying models. The ratio of $t\bar t$ production from light quarks over from gluons can be increased by as large as 100\%. We have also included our calculated best spin quantization axis to maximize the top quark polarization and spin-correlation effects in the appendix. We believe that those axes could be very useful to distinguish different new physics models eventually.

\subsection*{Acknowledgments} 
We would like to thank J. L. Hewett, J. Shelton and especially Thomas G. Rizzo for useful discussions and comments. SLAC is operated by Stanford University for the US Department of Energy under contract DE-AC02-76SF00515. Zhenyu Han is partially supported by the NSF under grants PHY-0804450.

%%%%%%%%%%%%%%%%%%%%%%%%%%%%%%%%
\appendix

\section{Formulas for top polarization and spin correlation}
\label{sec:formulas}

%%%%%%%%%%%%%%%%%%%%%%%%%%%%%%%%
\subsection{The axigluon case}
\label{sec:formulaaxigluon}

We assume that the axigluon has universal vector and axial couplings to the first two generations, $g_V^q$ and $g_A^q$, while the couplings to the top quark and bottom quark will be $g_V^t$ and $g_A^t$.  For convenience we have rescaled these couplings in terms of the strong coupling $g_s$. The tree-level differential cross section for $q \bar q \to t \bar t$ will be \cite{Hewett:1988xc,Ferrario:2009bz} 
\bea
\label{eqn:phenom}
\frac{d \hat{\sigma}^{q \bar q \to t \bar t}}{d \cos \theta^*} &=& \alpha_s^2 \frac{\pi \beta}{9 \hat{s}} \left[ \left(1 + 4m^2 + c^2\right) \left(1 -  \frac{2  g_V^q g_V^t  \hat{s}(M_{G^\prime}^2 - \hat{s}) }{(\hat{s}-M_{G^\prime}^2)^2+M_{G^\prime}^2 \Gamma_G^2}   
+  \frac{g_V^{t\,2}  (g_V^{q\,2} + g_A^{q\,2}) \hat{s}^2}{(\hat{s}-M_{G^\prime}^2)^2+M_{G^\prime}^2 \Gamma_G^2}  \right)   \right. \nonumber \\ 
& & + \left(1 - 4m^2 + c^2 \right) g_A^{t\,2}  (g_V^{q\,2} + g_A^{q\,2}) \frac{\hat{s}^2}{(\hat{s}-M_{G^\prime}^2)^2+M_{G^\prime}^2 \Gamma_G^2} \nonumber \\
& & - \left.  4 g_A^q g_A^t c \left(  \frac{\hat{s}(M_{G^\prime}^2-\hat{s}) }{(\hat{s}-M_{G^\prime}^2)^2+M_{G^\prime}^2 \Gamma_G^2}  -  2 g_V^q g_V^t \frac{\hat{s}^2}{(\hat{s}-M_{G^\prime}^2)^2+M_{G^\prime}^2 \Gamma_G^2} \right) \right]
\eea
where $m^2 = m_t^2/\hat{s}$,  $\beta = \sqrt{1-4m^2}$, $c = \beta\cos \theta^* $, and $\theta^*$ is the angle between the top quark and  the incoming quark in the center of mass frame.  The forward-backward asymmetry arises solely from the last line in this equation, so to obtain a positive asymmetry we must have $g_A^q g_A^t < 0$.

To study the top and anti-top quark polarizations and the spin correlations of top and anti-top quarks, we calculate the spin-density of the top quark pair productions, following Ref.~\cite{Bernreuther:1993hq} and Ref.~\cite{Uwer:2004vp}. The matrix element square is decomposed into different spin structures of top and anti-top
\beq
\frac{1}{2^2\,N_c^2}\sum_{spin (initial), color}|{\cal{M}}|^2 \,=\, {\rm Tr}\left[ \rho \cdot \frac{1}{2}(\mathbb{I}_2 + \hat{s}_t \cdot {\bf\sigma}) \otimes \frac{1}{2}(\mathbb{I}_2 + \hat{s}_{\bar t} \cdot {\bf\sigma}) \right]\,.
\eeq
Here, $\hat{s}_t$ ($\hat{s}_{\bar t}$) is the unit polarization (three-component) of the top (anti-top) quark in the rest frame of the top (anti-top) quark and $\sigma^i$ are Pauli matrices. The spin density function $\rho$ is a $4\times 4$ matrix in terms of the four-components of the spins of top and anti-top, which defined as 
\beqa
s^\mu_t &=& \left( \frac{ {{\bf k} \cdot \hat{s}_t} }{m_t}, \hat{s}_t +  \frac{{\bf k} ( {\bf k} \cdot \hat{s}_t)  }{m_t(m_t+E)}  \right) \,, \nonumber \\
s^\mu_{\bar t} &=& \left( - \frac{ {{\bf k} \cdot \hat{s}_{\bar t}} }{m_t}, \hat{s}_{\bar t} +  \frac{{\bf k} ( {\bf k} \cdot \hat{s}_{\bar t})  }{m_t(m_t+E)}  \right) \,.
\eeqa
Here, $E$ and ${\bf k}$  are the energy and the three-momentum of the top quark in the $t\bar{t}$ rest frame. Those two four-component spin vectors satisfy the Bjorken and Drell relations: $s_t^2=s^2_{\bar t} = -1$ and $s_t\cdot k_t = s_{\bar t} \cdot k_{\bar t} = 0$ where $k_t$ and $k_{\bar t}$  are the top and anti-top quarks four-momenta in the center-of-mass frame. To calculate the matrix element, the following relations of spinor outer products $u\bar{u}$ and $v\bar{v}$ are useful: $u(p, s) \bar{u} (p, s) = \frac{1}{2} (\slsh{p} + m)(1+\gamma_5 \slsh{s})$ and $v(p, s) \bar{v} (p, s) = \frac{1}{2} (\slsh{p} - m)(1+\gamma_5 \slsh{s})$ (see Ref.~\cite{Mahlon:1995zn} for more detailed discussions).  

The spin density matrix is decomposed into independent parts with different spin structures
\beq
\rho = A\, {\mathbb{I}_2}\otimes \mathbb{I}_2 + {\bf B}_t \cdot {\bf \sigma} \otimes \mathbb{I}_2 + {\bf B}_{\bar t} \cdot  \mathbb{I}_2 \otimes {\bf \sigma} + C_{ij} \sigma^i \otimes \sigma^j \,,
\eeq
whether ${\bf B}_t$ (${\bf B}_{\bar t}$) determines the top (anti-top) quark polarizations and $C_{ij}$ denotes the potential spin-correlations between top and anti-top, especially when ${\bf B}_t= {\bf B}_{\bar t}=0$. To determine the spin-quantization axis and without loss of generalities, we choose a coordinate with the unit vectors of the initial up quark and the final top quark as 
\beqa
\hat{\bf p} = (0, 0, 1)^T \,, \qquad \hat{\bf k} = (0, s_{\theta^*}, c_{\theta^*})^T\,,
\eeqa
with $s_{\theta^*}=\sin{\theta^*}$ and $c_{\theta^*}=\cos{\theta^*}$. One can also define one more unit vector perpendicular to both $\hat{\bf p}$ and $\hat{\bf k}$ as $\hat{\bf n} = \hat{\bf p} \times \hat{\bf k}/|\hat{\bf p} \times \hat{\bf k}|$. The matrices ${\bf B}_t$, ${\bf B}_{\bar t}$, $C$ can be further decomposed into
\beqa
{\bf B}_t &=& b_t^{\hat p}\, \hat{\bf p} + b_t^{\hat k}\, \hat{\bf k}  + b_t^{\hat n}\, \hat{\bf n}  \,,  \\
{\bf B}_{\bar t} &=& b_{\bar t}^{\hat p}\, \hat{\bf p} + b_{\bar t}^{\hat k}\, \hat{\bf k}  + b_{\bar t}^{\hat n}\, \hat{\bf n} \,, \\
C &= & c_0 \,{\mathbb{I}_3} + c_4\,\hat{\bf p} \otimes \hat{\bf p} + c_5\,\hat{\bf k} \otimes \hat{\bf k} + c_6\,(\hat{\bf k} \otimes \hat{\bf p} + \hat{\bf p} \otimes \hat{\bf k} ) \,.
\eeqa
Here, assuming CPT is a good symmetry, we have $b_t^{\hat n}=b_{\bar t}^{\hat n} = 0$. Furthermore, we will assume that CP is a good symmetry for the axigluon model, so one has $b_t^{\hat p} = b_{\bar t}^{\hat p}$ and $b_t^{\hat k} = b_{\bar t}^{\hat k}$. Other structures of the matrix $C$ are also forbidden by CP and CPT symmetries. Under P-symmetry transformation, one has $b_{t, \hat t}^{\hat p} \rightarrow  - b_{t, \hat t}^{\hat p}$ and $b_{t, \hat t}^{\hat k} \rightarrow  - b_{t, \hat t}^{\hat k}$. So, if parity-symmetry is a good symmetry, we anticipate ${\bf B}_t = {\bf B}_{\bar t}  =0$. 

In the QCD and for the production processes $q\bar{q} \rightarrow t \bar{t}$ at tree level, one has
\beqa
A^{\rm QCD} &=& \frac{1}{18} \left( 2 -\beta^2\,s^2_{\theta^*}    \right) \,, \qquad {\bf B}^{\rm QCD}_t = {\bf B}^{\rm QCD}_{\bar t}  =0 \,, \\
c^{\rm QCD}_0&=& -\frac{\beta^2 \,s^2_{\theta^*} }{18} \,,\qquad c^{\rm QCD}_4 = \frac{1}{9} \,, \\
c^{\rm QCD}_5&=& \frac{\beta^2 \left( 4m - \beta^2  s^2_{\theta^*} +2 \right)}{9 (2m + 1)^2} \,, \qquad
c^{\rm QCD}_6= \frac{ \left( 2m - 1 \right)  c_{\theta^*} }{9} \,.
\eeqa
An overall coefficient $g_s^4$  is understood. The interference terms of QCD and axigluon contributions have the following formulas
\beqa
A^{\rm int} &=& \frac{\hat{s}\left(\hat{s} - M^2_{G^\prime}\right)}{18 \left[ (\hat{s}  - M^2_{G^\prime})^2 + M^2_{G^\prime} \Gamma^2_{G^\prime}  \right]}\left[4 g_A^u g_A^t \beta \,c_{\theta^*} + 2 g_V^u g_V^t (2 - \beta^2 s^2_{\theta^*}  ) \right] \,, \\
b^{\rm int}_{\hat{p}} &=& \frac{\hat{s}\left(\hat{s} - M^2_{G^\prime}\right)}{18 \left[ (\hat{s}  - M^2_{G^\prime})^2 + M^2_{G^\prime} \Gamma^2_{G^\prime}  \right]}\left[4 m \, g_V^u g_A^t \beta \,c_{\theta^*}  + 8 m\,g_V^t g_A^u \right] \,, \\
b^{\rm int}_{\hat{k}} &=& \frac{\hat{s}\left(\hat{s} - M^2_{G^\prime}\right)}{18 \left[ (\hat{s}  - M^2_{G^\prime})^2 + M^2_{G^\prime} \Gamma^2_{G^\prime}  \right]}\left[ 2 g_V^u g_A^t \,\beta - 2c_{\theta^*}(2m -1) (g_V^u g_A^t \beta \,c_{\theta^*} + 2 \,g_V^t g_A^u) \right] \,,
\eeqa
and 
\beqa
c^{\rm int}_0 &=&  \frac{\hat{s}\left(\hat{s} - M^2_{G^\prime}\right)}{18 \left[ (\hat{s}  - M^2_{G^\prime})^2 + M^2_{G^\prime} \Gamma^2_{G^\prime}  \right]}\left[ - 2 g_V^u g_V^t \beta^2 \,s^2_{\theta^*} \right] \,, \\
c^{\rm int}_4 &=&  \frac{\hat{s}\left(\hat{s} - M^2_{G^\prime}\right)}{18 \left[ (\hat{s}  - M^2_{G^\prime})^2 + M^2_{G^\prime} \Gamma^2_{G^\prime}  \right]}\left[  4 g_V^u g_V^t \right] \,, \\
c^{\rm int}_5 &=&  \frac{\hat{s}\left(\hat{s} - M^2_{G^\prime}\right)}{18 \left[ (\hat{s}  - M^2_{G^\prime})^2 + M^2_{G^\prime} \Gamma^2_{G^\prime}  \right]}  \frac{ \left[ 4 g_A^u g_A^t  (2m+1) \beta^3\, c_{\theta^*}
+ 4\beta^2\,g_V^t g_V^u (4m - \beta^2 s^2_{\theta^*} +2 ) \right] }{(2m+1)^2} \,, \\
c^{\rm int}_6 &=&  \frac{\hat{s}\left(\hat{s} - M^2_{G^\prime}\right)}{18 \left[ (\hat{s}  - M^2_{G^\prime})^2 + M^2_{G^\prime} \Gamma^2_{G^\prime}  \right]}   \left[ 4 g_A^u g_A^t  m\, \beta
+ 4\,g_V^t g_V^u (2m - 1)c_{\theta^*} \right] \,,
\eeqa
The new-physics-only part has
\beqa
A^{\rm new} &=& \frac{\hat{s}^2\left\{ \frac{1}{2} (g_V^{u\,2}+g_A^{u\,2}) \left[\beta^2 (g_V^{t\,2}+g_A^{t\,2})  c_{2\theta^*}+ 3 \beta^2 g_A^{t\,2} + g_V^{t\,2} (4 - \beta^2) \right] + 8 g_V^u g_V^t g_A^u g_A^t \beta\,c_{\theta^*}  
\right\}}{18 \left[ (\hat{s}  - M^2_{G^\prime})^2 + M^2_{G^\prime} \Gamma^2_{G^\prime}  \right]} \,, \\
b^{\rm new}_{\hat{p}} &=& \frac{\hat{s}^2
 \left[ 4 \,m\,g_V^t\,g_A^t \,\beta  (g_A^{u\,2}  + g_V^{u\,2}) c_{\theta^*}  + 8\,m\,g_A^u g_V^u g_V^{t\,2}   \right]
}{18 \left[ (\hat{s}  - M^2_{G^\prime})^2 + M^2_{G^\prime} \Gamma^2_{G^\prime}  \right]} \,, \\
b^{\rm new}_{\hat{k}} &=& \frac{-\,2\, \beta\, \hat{s}^2
}{18 (2m+1)^2\, s_{\theta^*}\,\left[ (\hat{s}  - M^2_{G^\prime})^2 + M^2_{G^\prime} \Gamma^2_{G^\prime}  \right]} 
\left\{  g_A^u g_V^u \beta\,s_{2\theta^*} \left[ g_A^{t\,2} (-4m + \beta^2 -2) - g_V^{t\,2} (2m + 1) \right] \right.  \nonumber \\
&& \left.
- \frac{1}{2} g_A^t g_V^t (g_A^{u\, 2} +   g_V^{u\,2} ) s_{\theta^*} \left[ (2m+1)\beta^2 c_{2\theta^*} + (2m-1) \beta^2 + 4 (2m+1)   \right] 
\right\}   \,.
\eeqa
and 
\beqa
c_0^{\rm new} &=& \frac{\hat{s}^2
 \left[ \beta^2 (g_A^{t\,2} - g_V^{t\,2})(g_A^{u\,2} + g_V^{u\,2}) s^2_{\theta^*}
   \right]
}{18 \left[ (\hat{s}  - M^2_{G^\prime})^2 + M^2_{G^\prime} \Gamma^2_{G^\prime}  \right]} \,, \\
c_4^{\rm new} &=& \frac{\hat{s}^2
 \left[ 2 (g_A^{u\,2} + g_V^{u\,2}) (g_V^{t\,2}  - \beta^2 \,g_A^{t\,2}  )
   \right]
}{18 \left[ (\hat{s}  - M^2_{G^\prime})^2 + M^2_{G^\prime} \Gamma^2_{G^\prime}  \right]} \,, \\
c_5^{\rm new} &=& \frac{\hat{s}^2
 \left[ 8 \beta^3 g_A^t g_V^t g_A^u g_V^u (2m +1 ) c_{\beta^*} + \beta^2 g_V^{t\,2} (g_A^{u\,2} + g_V^{u\,2}) 
 (8 m - 2 \beta^2 s^2_{\theta^*}+ 4)   \right]
}{18 \left[ (\hat{s}  - M^2_{G^\prime})^2 + M^2_{G^\prime} \Gamma^2_{G^\prime}  \right] \,(2m+1)^2} \,, \\
c_6^{\rm new} &=& \frac{\hat{s}^2
 \left[ 2 (g_A^{u\,2} + g_V^{u\,2} ) (g_A^{t\,2} \beta^2 + g_V^{t\,2} (2m-1)) c_{\theta^*} + 8 g_A^t g_A^u g_V^t g_V^u m\, \beta  \right]
}{18 \left[ (\hat{s}  - M^2_{G^\prime})^2 + M^2_{G^\prime} \Gamma^2_{G^\prime}  \right]} \,, 
\eeqa

From Eq.~(\ref{eqn:phenom}) and to maximize the effects of increasing $A^{t\bar t}_{FB}$, we choose $g_V^t = g_V^u =0$. For this choice of parameters, all $b$'s are zero and there is no polarizations of top and anti-top quarks at the tree level. The summations of $c$'s are
\beqa
c_0 & =&  -\frac{\beta^2 \,s^2_{\theta^*} }{18}  + \frac{\hat{s}^2
 \beta^2 g_A^{t\,2} g_A^{u\,2} s^2_{\theta^*}  
}{18 \left[ (\hat{s}  - M^2_{G^\prime})^2 + M^2_{G^\prime} \Gamma^2_{G^\prime}  \right]}  \,, \\
c_4 & =&  \frac{1}{9}  -   \frac{ \hat{s}^2
 \beta^2  \,g_A^{u\,2}  \,g_A^{t\,2}
}{9 \left[ (\hat{s}  - M^2_{G^\prime})^2 + M^2_{G^\prime} \Gamma^2_{G^\prime}  \right]} \,, \\
c_5 & =& \frac{\beta^2 \left( 4m - \beta^2  s^2_{\theta^*} +2 \right)}{9 (2m + 1)^2} 
+  \frac{\hat{s}\left(\hat{s} - M^2_{G^\prime}\right) 2 g_A^u g_A^t  \beta^3\, c_{\theta^*}}{9 \left[ (\hat{s}  - M^2_{G^\prime})^2 + M^2_{G^\prime} \Gamma^2_{G^\prime}  \right] (2m+1) }  \,, \\
c_6 & =&\frac{\left( 2m - 1 \right)  c_{\theta^*} }{9} +
\frac{\hat{s}\left(\hat{s} - M^2_{G^\prime}\right) 2 g_A^u g_A^t  m\, \beta }{9 \left[ (\hat{s}  - M^2_{G^\prime})^2 + M^2_{G^\prime} \Gamma^2_{G^\prime}  \right]}   
+
\frac{\hat{s}^2\,\beta^2
g_A^{u\,2} g_A^{t\,2}  c_{\theta^*} 
}{9 \left[ (\hat{s}  - M^2_{G^\prime})^2 + M^2_{G^\prime} \Gamma^2_{G^\prime}  \right]} 
 \,,
\eeqa

To determine the spin axis to maximize the spin correlations between top and anti-top quarks, one needs to diagonalize the $C_{ij}$ matrix 
\beq
C = 
\left( \begin{array}{ccc}
c_0 & 0 & 0 \\
0 & c_0+ c_5 s^2_{\theta^*} & c_6 s_{\theta^*} + c_5 c_{\theta^*}  s_{\theta^*}   \\
0 & c_6 s_{\theta^*} + c_5 c_{\theta^*}  s_{\theta^*}  & c_0 + c_4 + c_5 c^2_{\theta^*} + 2 c_6  c_{\theta^*} \end{array} \right) \,,
\eeq
and find the eigenvector corresponding to the largest eigenvalue. 

%%%%%%%%%%%%%%%%%%%%%%%%%%%%%%%%
\subsection{Contact Operator}
\label{sec:appxContact}
We first analyze the contact operator case. The relevant operator can be written as $\bar{u} \gamma_\mu\,\tau^a\, \gamma^5 u \,\bar{t} \gamma^\mu \gamma_5\,\tau^a\, t /\Lambda^2$ with $\tau^a$ $SU(3)_{\rm QCD}$ generators. To match the formulas for the axigluon model, we identify
\beq
\frac{g^u_A g^t_A}{M^2_{G^\prime}} = \frac{\xi}{\Lambda^2}\,,
\eeq
and $\Gamma_{G^\prime} = 0$. Here, $\xi = +1 (-1)$ for $g^u_A g^t_A >0 (<0)$. The positive $A_{FB}$ observed at Tevatron prefers to have $\xi = -1$. The spin-correlation matrix has
\beqa
c_0 & =&  -\frac{\beta^2 \,s^2_{\theta^*} }{18}  + \frac{\hat{s}^2
 \beta^2 s^2_{\theta^*}  
}{18 \Lambda^4}  \,, \\
c_4 & =&  \frac{1}{9}  -   \frac{\hat{s}^2
 \beta^2  
}{9 \Lambda^4} \,, \\
c_5 & =& \frac{\beta^2 \left( 4m - \beta^2  s^2_{\theta^*} +2 \right)}{9 (2m + 1)^2} 
-  \frac{2\xi\,\hat{s} \beta^3\, c_{\theta^*}}{9 \Lambda^2 (2m+1) }  \,, \\
c_6 & =&\frac{\left( 2m - 1 \right)  c_{\theta^*} }{9} -
\frac{2\xi\,\hat{s}    m\, \beta }{9 \Lambda^2 }   
+
\frac{\hat{s}^2\,\beta^2 c_{\theta^*} 
}{9\Lambda^4} 
 \,,
\eeqa
Diagonalize the $C$ matrix, one has the three eigenvalues as
\beqa
\pm \frac{  \beta^2 (1- c^2_{\theta^*} )}{18} (1 - \frac{\hat{s}^2}{\Lambda^4} )\,,
\quad
\frac{2 - \beta^2 (1- c^2_{\theta^*} )}{18} - \frac{2\xi\hat{s} \,c_{\theta^*}  \, \beta }{9\,\Lambda^2}
+ \frac{\beta^2 \hat{s}^2 (1+c^2_{\theta^*} )}{18 \,\Lambda^4} \,.
\eeqa
One can easily check that the last eigenvalue is always the largest one, independent of $\hat{s}$ and $\Lambda$. So, this provides us a fixed axis to quantize the spins of top and anti-top quarks. Simplifying the eigenvector corresponding to the largest eigenvalue, we have the following best spin quantization axis for spin correlation
\beqa
{\bf{e}}^q \propto \hat{p} +  \left[ c_{\theta^*} (\gamma-1) - \frac{\xi\,\beta\gamma \hat{s}}{\Lambda^2}  \right]\,\hat{k} \,, 
%\label{eq:cutoffaxis}
\eeqa
with $\gamma = 1/\sqrt{1-\beta^2}$. For the case that $\Lambda \gg \hat{s}$ and the new physics is decoupled, 
we recover the formula for the axis maximizing spin-correlation in the SM~\cite{Mahlon:1997uc}.

Coming back to the axigluon model and neglecting the resonance width, we obtain the best quantization axis to be
\beq
{\bf{e}}^q \propto \hat{p} +  \left[ c_{\theta^*} (\gamma-1) - \frac{\beta\gamma \hat{s}\,g_A^u\,g_A^t}{M_{G^\prime}^2 - \hat{s}}  \right]\,\hat{k} \,.
\eeq
%

%%%%%%%%%%%%%%%%%%%%%%%%%%%%%%%%
\subsection{Flavor violating $W^\prime$ models}
\label{sec:formulawprime}

Considering the following interactions,
\beq
\bar{t} (g_V \gamma^\mu + g_A \gamma^\mu\gamma^5) d \,W^{\prime +}_\mu \,+\,h.c.,
 \eeq
for the interference term, we have
\begin{eqnarray}
A^{\text{int}}&=&-\frac{s \left(g_A^2+g_V^2\right) \left(\frac{1}{2} m^2 \left((-1+c)^2+4 m^2\right) s+\left((1+c)^2+4 m^2\right) M_{W'}^2\right)}{9 M_{W'}^2 \left(-t+M_{W'}^2\right)},\\
b^{\text{int}}_{\hat p}&=&\frac{4 m s g_A g_V \left(\frac{1}{2} (-2+c) m^2 s-(2+c) M_{W'}^2\right)}{9 M_{W'}^2 \left(-t+M_{W'}^2\right)},\\
b^{\text{int}}_{\hat k}&=&\frac{2 s \beta  g_A g_V \left(\frac{1}{2} m^2 \left((-1+c)^2+2 m\right) s-\left((1+c)^2+2 m\right) M_{W'}^2\right)}{9 (1+2 m) M_{W'}^2 \left(-t+M_{W'}^2\right)},\\
c_0^{\text{int}}&=&-\frac{\left(-1+c^2+4 m^2\right) s \left(g_A^2+g_V^2\right) \left(\frac{m^2 s}{2}+M_{W'}^2\right)}{9 M_{W'}^2 \left(-t+M_{W'}^2\right)},\\
c_4^{\text{int}}&=&\frac{2 s \left(g_A^2+g_V^2\right) \left(\frac{m^2 s}{2}+M_{W'}^2\right)}{9 M_{W'}^2 \left(t-M_{W'}^2\right)},\\
c_5^{\text{int}}&=&\frac{c s \beta ^2 \left(g_A^2+g_V^2\right) \left(m^2 s-2 M_{W'}^2\right)}{9 (1+2 m) M_{W'}^2 \left(-t+M_{W'}^2\right)}+\nonumber\\
&&\frac{s \beta ^2 \left(32 m^3+48 m^4+4 m^2 \left(1+\beta^2+2z^2\beta^2\right)\right) \left(g_A^2+g_V^2\right) \left(m^2 s+2 M_{W'}^2\right)}{72 m^2 (1+2 m)^2 M_{W'}^2 \left(t-M_{W'}^2\right)},\\
c_6^{\text{int}}&=&\frac{2 s \beta  \left(g_A^2+g_V^2\right) \left(\frac{1}{2} m^2 \left(c+m+2 m^2\right) s-\left(-c+m+2 m^2\right) M_{W'}^2\right)}{9 (1+2 m) M_{W'}^2 \left(-t+M_{W'}^2\right)}.
\end{eqnarray}

For the new physics squared diagram, 
%{\footnotesize
\begin{eqnarray}
A^{\text{new}}&=&\frac{s^2 \left(\left(5-8 m^2+c (2+z \beta )\right) (g_V^4+g_A^4)+2 \left(-1+8 m^2+3 c (2+z \beta )\right) g_A^2 g_V^2\right)}{8 \left(t-M_{W'}^2\right){}^2}+\nonumber\\
&&\frac{m^4 s^3 \left(g_A^2+g_V^2\right){}^2 \left((-1+c)^2 s+16 M_{W'}^2\right)}{16 M_{W'}^4 \left(t-M_{W'}^2\right){}^2},\\
b^{\text{new}}_{\hat p} &=& -\frac{m s^2 g_A g_V \left(g_A^2+g_V^2\right) \left(\frac{m^2 s}{2}+M_{W'}^2\right) \left(\frac{1}{2} (-1+c) m^2 s-(1+c) M_{W'}^2\right)}{M_{W'}^4 \left(-t+M_{W'}^2\right){}^2},\\
b^{\text{new}}_{\hat k}&=&\frac{s^2 \beta  g_A g_V \left(g_A^2+g_V^2\right) \left(\frac{1}{4} (-1+c) m^4 (1-c+2 m) s^2-2 c m^3 s M_{W'}^2+(1+c) (1+c+2 m) M_{W'}^4\right)}{2 (1+2 m) M_{W'}^4 \left(-t+M_{W'}^2\right){}^2}.  \nonumber \\
\end{eqnarray}

\begin{eqnarray}
c_0^{\text{new}}&=&-\frac{m^2 s^2 \left(g_A^2-g_V^2\right){}^2}{\left(t-M_{W'}^2\right){}^2}+\nonumber\\
&&\frac{m^2 s^3 \left(2 \left((-1+c) (1+3 c)+8 m^2\right) g_A^2 g_V^2+\left(-3+c (2+c)+8 m^2\right) \left(g_A^4+g_V^4\right)\right)}{8 \left(-t M_{W'}+M_{W'}^3\right){}^2},\\
c_4^{\text{new}}&=&\frac{m^2 s^2 \left(2 m^4 s^2 g_A^2 g_V^2-\frac{1}{2} \left(-1+2 m^2\right) s \left(g_A^4+6 g_A^2 g_V^2+g_V^4\right) M_{W'}^2+\left(g_A^4+6 g_A^2 g_V^2+g_V^4\right) M_{W'}^4\right)}{2 M_{W'}^4 \left(-t+M_{W'}^2\right){}^2},\\
c_5^{\text{new}}&=&\frac{1}{8 (1+2 m)^2 M_{W'}^4 \left(-t+M_{W'}^2\right){}^2}s^2 \beta ^2 \left(2 m^4 (1-c+2 m)^2 s^2 g_A^2 g_V^2+\right.\nonumber\\
&&\frac{1}{4} s M_{W'}^2\left(2 \left(16 (1+c) m^3+48 m^4+(-1+c) (1+3 c) \left(-1+\beta ^2\right)+8 m^2 \left(-1+3 c^2+\beta ^2\right)\right) g_A^2 g_V^2+\right.\nonumber\\
&&\left.\left(-16 (-3+c) m^3+80 m^4+(-1+c) (3+c) \left(-1+\beta ^2\right)+8 m^2 \left(-1+\left(1+z^2\right) \beta ^2\right)\right) \left(g_A^4+g_V^4\right)\right) +\nonumber\\
&& M_{W'}^4\left(2 \left(5+4 (7+3 c) m+36 m^2+\beta  (3 (2+c) z+2 \beta )\right) g_A^2 g_V^2\right.\nonumber\\
&&\left.\left.+\left(-1+2 c+4 c m-4 m (3+5 m)+\left(-2+z^2\right) \beta ^2\right) \left(g_A^4+g_V^4\right)\right)\right),\\
c_6^{\text{new}}&=&\frac{m s^2 \beta }{4 (1+2 m) M_{W'}^4 \left(-t+M_{W'}^2\right){}^2}\left(-2 m^4 (1-c+2 m) s^2 g_A^2 g_V^2+\right.\nonumber\\
&&\left.m s \left(-m (1+2 m) \left(g_A^2-g_V^2\right){}^2-c (1+m) \left(g_A^4+6 g_A^2 g_V^2+g_V^4\right)\right) M_{W'}^2+\right.\nonumber\\
&&\left.(1+c+2 m) \left(g_A^4+6 g_A^2 g_V^2+g_V^4\right) M_{W'}^4\right).
\end{eqnarray}
%}

%%%%%%%%%%%%%%%%%%%%%%%%%%%%%%%%%%%
%\bibliography{topFBSpin}

\begin{thebibliography}{10}

\bibitem{AttCDF}
{\bf The CDF Collaboration} Collaboration, T.~Aaltonen {\em et.~al.}, {\it
  {Evidence for a Mass Dependent Forward-Backward Asymmetry in Top Quark Pair
  Production}},  \href{http://xxx.lanl.gov/abs/1101.0034}{{\tt
  arXiv:1101.0034}}.

\bibitem{Abazov:2007qb}
{\bf D0} Collaboration, V.~M. Abazov {\em et.~al.}, {\it {First measurement of
  the forward-backward charge asymmetry in top quark pair production}},  {\em
  Phys. Rev. Lett.} {\bf 100} (2008) 142002,
  [\href{http://xxx.lanl.gov/abs/0712.0851}{{\tt arXiv:0712.0851}}].

\bibitem{CDFAFBdilepton}
{\bf CDF Collaboration} Collaboration, C.~Collaboration, {\it {Measurement of
  the Forward Backward Asymmetry in Top Pair Production in the Dilepton Decay
  Channel using 5.1 $fb^{-1}$}},  {\em CDF Note 10436} (2011).

\bibitem{Ferrario:2009bz}
P.~Ferrario and G.~Rodrigo, {\it {Constraining heavy colored resonances from
  top-antitop quark events}},  {\em Phys. Rev.} {\bf D80} (2009) 051701,
  [\href{http://xxx.lanl.gov/abs/0906.5541}{{\tt arXiv:0906.5541}}].

\bibitem{Frampton:2009rk}
P.~H. Frampton, J.~Shu, and K.~Wang, {\it {Axigluon as Possible Explanation for
  $p\bar{p} \to t\bar{t}$ Forward-Backward Asymmetry}},  {\em Phys. Lett.} {\bf
  B683} (2010) 294--297, [\href{http://xxx.lanl.gov/abs/0911.2955}{{\tt
  arXiv:0911.2955}}].

\bibitem{Shu:2009xf}
J.~Shu, T.~M.~P. Tait, and K.~Wang, {\it {Explorations of the Top Quark
  Forward-Backward Asymmetry at the Tevatron}},  {\em Phys. Rev.} {\bf D81}
  (2010) 034012, [\href{http://xxx.lanl.gov/abs/0911.3237}{{\tt
  arXiv:0911.3237}}].

\bibitem{Cao:2010zb}
Q.-H. Cao, D.~McKeen, J.~L. Rosner, G.~Shaughnessy, and C.~E.~M. Wagner, {\it
  {Forward-Backward Asymmetry of Top Quark Pair Production}},  {\em Phys. Rev.}
  {\bf D81} (2010) 114004, [\href{http://xxx.lanl.gov/abs/1003.3461}{{\tt
  arXiv:1003.3461}}].

\bibitem{Chivukula:2010fk}
R.~S. Chivukula, E.~H. Simmons, and C.~P. Yuan, {\it {Axigluons cannot explain
  the observed top quark forward- backward asymmetry}},  {\em Phys. Rev.} {\bf
  D82} (2010) 094009, [\href{http://xxx.lanl.gov/abs/1007.0260}{{\tt
  arXiv:1007.0260}}].
  
    \bibitem{Bai:2011ed}
Y.~Bai, J.~L. Hewett, J.~Kaplan, and T.~G. Rizzo, {\it {LHC Predictions from a
  Tevatron Anomaly in the Top Quark Forward-Backward Asymmetry}},  {\em JHEP}
  {\bf 03} (2011) 003, [\href{http://xxx.lanl.gov/abs/1101.5203}{{\tt
  arXiv:1101.5203}}].

\bibitem{Berger:2011ua}
E.~L. Berger, Q.-H. Cao, C.-R. Chen, C.~S. Li, and H.~Zhang, {\it {Top Quark
  Forward-Backward Asymmetry and Same-Sign Top Quark Pairs}},  {\em
  Phys.Rev.Lett.} {\bf 106} (2011) 201801,
  [\href{http://xxx.lanl.gov/abs/1101.5625}{{\tt arXiv:1101.5625}}].

\bibitem{Barger:2011ih}
V.~Barger, W.-Y. Keung, and C.-T. Yu, {\it {Tevatron Asymmetry of Tops in a
  W',Z' Model}},  {\em Phys.Lett.} {\bf B698} (2011) 243--250,
  [\href{http://xxx.lanl.gov/abs/1102.0279}{{\tt arXiv:1102.0279}}].

\bibitem{Bhattacherjee:2011nr}
B.~Bhattacherjee, S.~S. Biswal, and D.~Ghosh, {\it {Top quark forward-backward
  asymmetry at Tevatron and its implications at the LHC}},  {\em Phys.Rev.}
  {\bf D83} (2011) 091501, [\href{http://xxx.lanl.gov/abs/1102.0545}{{\tt
  arXiv:1102.0545}}].

\bibitem{Blum:2011up}
K.~Blum, C.~Delaunay, O.~Gedalia, Y.~Hochberg, S.~J. Lee, {\em et.~al.}, {\it
  {Implications of the CDF $t\bar{t}$ Forward-Backward Asymmetry for Boosted
  Top Physics}},  \href{http://xxx.lanl.gov/abs/1102.3133}{{\tt
  arXiv:1102.3133}}.

\bibitem{Grinstein:2011yv}
B.~Grinstein, A.~L. Kagan, M.~Trott, and J.~Zupan, {\it {Forward-backward
  asymmetry in t anti-t production from flavour symmetries}},
  \href{http://xxx.lanl.gov/abs/1102.3374}{{\tt arXiv:1102.3374}}.

\bibitem{Ligeti:2011vt}
Z.~Ligeti, M.~Schmaltz, and G.~M. Tavares, {\it {Explaining the t tbar
  forward-backward asymmetry without dijet or flavor anomalies}},
  \href{http://xxx.lanl.gov/abs/1103.2757}{{\tt arXiv:1103.2757}}.

\bibitem{Gresham:2011pa}
M.~I. Gresham, I.-W. Kim, and K.~M. Zurek, {\it {On Models of New Physics for
  the Tevatron Top $A_{FB}$}},  \href{http://xxx.lanl.gov/abs/1103.3501}{{\tt
  arXiv:1103.3501}}.
  
\bibitem{Nelson:2011us}
  A.~E.~Nelson, T.~Okui and T.~S.~Roy, {\it {A unified, flavor symmetric explanation for the t-tbar asymmetry and Wjj excess at CDF}},  \href{http://xxx.lanl.gov/abs/1104.2030}{{\tt
  arXiv:1104.2030}}.

\bibitem{Barcelo:2011fw}
R.~Barcelo, A.~Carmona, M.~Masip, and J.~Santiago, {\it {Gluon excitations in t
  tbar production at hadron colliders}},
  \href{http://xxx.lanl.gov/abs/1105.3333}{{\tt arXiv:1105.3333}}.

\bibitem{Haisch:2011up}
U.~Haisch and S.~Westhoff, {\it {Massive Color-Octet Bosons: Bounds on Effects
  in Top-Quark Pair Production}},
  \href{http://xxx.lanl.gov/abs/1106.0529}{{\tt arXiv:1106.0529}}.

\bibitem{Cui:2011xy}
Y.~Cui, Z.~Han, and M.~D. Schwartz, {\it {Top condensation as a motivated
  explanation of the top forward-backward asymmetry}},
  \href{http://xxx.lanl.gov/abs/1106.3086}{{\tt arXiv:1106.3086}}.

\bibitem{Kuhn:1998jr}
J.~H. Kuhn and G.~Rodrigo, {\it {Charge asymmetry in hadroproduction of heavy
  quarks}},  {\em Phys. Rev. Lett.} {\bf 81} (1998) 49--52,
  [\href{http://xxx.lanl.gov/abs/hep-ph/9802268}{{\tt hep-ph/9802268}}].

\bibitem{Kuhn:1998kw}
J.~H. Kuhn and G.~Rodrigo, {\it {Charge asymmetry of heavy quarks at hadron
  colliders}},  {\em Phys. Rev.} {\bf D59} (1999) 054017,
  [\href{http://xxx.lanl.gov/abs/hep-ph/9807420}{{\tt hep-ph/9807420}}].

\bibitem{Antunano:2007da}
O.~Antunano, J.~H. Kuhn, and G.~Rodrigo, {\it {Top quarks, axigluons and charge
  asymmetries at hadron colliders}},  {\em Phys. Rev.} {\bf D77} (2008) 014003,
  [\href{http://xxx.lanl.gov/abs/0709.1652}{{\tt arXiv:0709.1652}}].

\bibitem{Wang:2010tg}
Y.-k. Wang, B.~Xiao, and S.-h. Zhu, {\it {One-side forward-backward asymmetry
  at the LHC}},  {\em Phys. Rev.} {\bf D83} (2011) 015002,
  [\href{http://xxx.lanl.gov/abs/1011.1428}{{\tt arXiv:1011.1428}}].

\bibitem{Xiao:2011kp}
B.~Xiao, Y.-K. Wang, Z.-Q. Zhou, and S.-h. Zhu, {\it {Edge Charge Asymmetry in
  Top Pair Production at the LHC}},  {\em Phys. Rev.} {\bf D83} (2011) 057503,
  [\href{http://xxx.lanl.gov/abs/1101.2507}{{\tt arXiv:1101.2507}}].

\bibitem{AguilarSaavedra:2011vw}
J.~A. Aguilar-Saavedra and M.~Perez-Victoria, {\it {Probing the Tevatron t tbar
  asymmetry at LHC}},  {\em JHEP} {\bf 05} (2011) 034,
  [\href{http://xxx.lanl.gov/abs/1103.2765}{{\tt arXiv:1103.2765}}].

\bibitem{Hewett:2011wz}
J.~L. Hewett, J.~Shelton, M.~Spannowsky, T.~M.~P. Tait, and M.~Takeuchi, {\it
  {$A^t_{FB}$ Meets LHC}},  \href{http://xxx.lanl.gov/abs/1103.4618}{{\tt
  arXiv:1103.4618}}.

\bibitem{Jezabek:1994zv}
M.~Jezabek and J.~H. Kuhn, {\it {V-A tests through leptons from polarized top
  quarks}},  {\em Phys. Lett.} {\bf B329} (1994) 317--324,
  [\href{http://xxx.lanl.gov/abs/hep-ph/9403366}{{\tt hep-ph/9403366}}].

\bibitem{Cao:2010nw}
J.~Cao, L.~Wu, and J.~M. Yang, {\it {New physics effects on top quark spin
  correlation and polarization at the LHC: a comparative study in different
  models}},  {\em Phys. Rev.} {\bf D83} (2011) 034024,
  [\href{http://xxx.lanl.gov/abs/1011.5564}{{\tt arXiv:1011.5564}}].

\bibitem{Jung:2010yn}
D.-W. Jung, P.~Ko, and J.~S. Lee, {\it {Longitudinal top polarization as a
  probe of a possible origin of forward-backward asymmetry of the top quark at
  the Tevatron}},  \href{http://xxx.lanl.gov/abs/1011.5976}{{\tt
  arXiv:1011.5976}}.

\bibitem{Choudhury:2010cd}
D.~Choudhury, R.~M. Godbole, S.~D. Rindani, and P.~Saha, {\it {Top
  polarization, forward-backward asymmetry and new physics}},
  \href{http://xxx.lanl.gov/abs/1012.4750}{{\tt arXiv:1012.4750}}.

\bibitem{Krohn:2011tw}
D.~Krohn, T.~Liu, J.~Shelton, and L.-T. Wang, {\it {A Polarized View of the Top
  Asymmetry}},  \href{http://xxx.lanl.gov/abs/1105.3743}{{\tt
  arXiv:1105.3743}}.

\bibitem{Mahlon:1995zn}
G.~Mahlon and S.~J. Parke, {\it {Angular correlations in top quark pair
  production and decay at hadron colliders}},  {\em Phys. Rev.} {\bf D53}
  (1996) 4886--4896, [\href{http://xxx.lanl.gov/abs/hep-ph/9512264}{{\tt
  hep-ph/9512264}}].

\bibitem{Stelzer:1995gc}
T.~Stelzer and S.~Willenbrock, {\it {Spin correlation in top quark production
  at hadron colliders}},  {\em Phys. Lett.} {\bf B374} (1996) 169--172,
  [\href{http://xxx.lanl.gov/abs/hep-ph/9512292}{{\tt hep-ph/9512292}}].

\bibitem{Mahlon:1997uc}
G.~Mahlon and S.~J. Parke, {\it {Maximizing spin correlations in top quark pair
  production at the Tevatron}},  {\em Phys. Lett.} {\bf B411} (1997) 173--179,
  [\href{http://xxx.lanl.gov/abs/hep-ph/9706304}{{\tt hep-ph/9706304}}].

\bibitem{Arai:2007ts}
M.~Arai, N.~Okada, K.~Smolek, and V.~Simak, {\it {Top quark spin correlations
  in the Randall-Sundrum model at the CERN Large Hadron Collider}},  {\em Phys.
  Rev.} {\bf D75} (2007) 095008,
  [\href{http://xxx.lanl.gov/abs/hep-ph/0701155}{{\tt hep-ph/0701155}}].

\bibitem{Frederix:2007gi}
R.~Frederix and F.~Maltoni, {\it {Top pair invariant mass distribution: a
  window on new physics}},  {\em JHEP} {\bf 01} (2009) 047,
  [\href{http://xxx.lanl.gov/abs/0712.2355}{{\tt arXiv:0712.2355}}].

\bibitem{Bai:2008sk}
Y.~Bai and Z.~Han, {\it {Top-antitop and Top-top Resonances in the Dilepton
  Channel at the CERN LHC}},  {\em JHEP} {\bf 04} (2009) 056,
  [\href{http://xxx.lanl.gov/abs/0809.4487}{{\tt arXiv:0809.4487}}].

\bibitem{Baumgart:2011wk}
M.~Baumgart and B.~Tweedie, {\it {Discriminating Top-Antitop Resonances using
  Azimuthal Decay Correlations}},
  \href{http://xxx.lanl.gov/abs/1104.2043}{{\tt arXiv:1104.2043}}.

\bibitem{Collaboration:2011dk}
C.~Collaboration, {\it {Search for Same-Sign Top-Quark Pair Production at
  sqrt(s) = 7 TeV and Limits on Flavour Changing Neutral Currents in the Top
  Sector}},  \href{http://xxx.lanl.gov/abs/1106.2142}{{\tt arXiv:1106.2142}}.

\bibitem{Hewett:1988xc}
J.~L. Hewett and T.~G. Rizzo, {\it {Low-Energy Phenomenology of Superstring
  Inspired E(6) Models}},  {\em Phys. Rept.} {\bf 183} (1989) 193.

\bibitem{Cheung:2009ch}
K.~Cheung, W.-Y. Keung, and T.-C. Yuan, {\it {Top Quark Forward-Backward
  Asymmetry}},  {\em Phys.Lett.} {\bf B682} (2009) 287--290,
  [\href{http://xxx.lanl.gov/abs/0908.2589}{{\tt arXiv:0908.2589}}].

\bibitem{Atlasttbar}
{\bf Atlas Collaboration} Collaboration, A.~Collaboration, {\it {A Search for
  $t\bar{t}$ Resonances in the Lepton Plus Jets Channel using 200 pb$^{-1}$ of
  $pp$ Collisions at $\sqrt{s} = 7$ TeV}},  {\em ATLAS-CONF-2011-087} (2011).

\bibitem{Alwall:2011uj}
J.~Alwall, M.~Herquet, F.~Maltoni, O.~Mattelaer, and T.~Stelzer, {\it {MadGraph
  5 : Going Beyond}},  \href{http://xxx.lanl.gov/abs/1106.0522}{{\tt
  arXiv:1106.0522}}.

\bibitem{Aaltonen:2010jr}
{\bf CDF} Collaboration, T.~Aaltonen {\em et.~al.}, {\it {Observation of Single
  Top Quark Production and Measurement of |Vtb| with CDF}},  {\em Phys. Rev.}
  {\bf D82} (2010) 112005, [\href{http://xxx.lanl.gov/abs/1004.1181}{{\tt
  arXiv:1004.1181}}].

\bibitem{Bernreuther:1993hq}
W.~Bernreuther and A.~Brandenburg, {\it {Tracing CP violation in the production
  of top quark pairs by multiple TeV proton proton collisions}},  {\em Phys.
  Rev.} {\bf D49} (1994) 4481--4492,
  [\href{http://xxx.lanl.gov/abs/hep-ph/9312210}{{\tt hep-ph/9312210}}].

\bibitem{Uwer:2004vp}
P.~Uwer, {\it {Maximizing the spin correlation of top quark pairs produced at
  the LHC}},  {\em Phys. Lett.} {\bf B609} (2005) 271--276,
  [\href{http://xxx.lanl.gov/abs/hep-ph/0412097}{{\tt hep-ph/0412097}}].

\end{thebibliography}
%\bibliographystyle{JHEP}
\providecommand{\href}[2]{#2}\begingroup\raggedright\endgroup

%%%%%%%%%%%%%%%%%%%%%%%%%%%%%%%%%%%
\end{document}